\documentclass[openacc]{rsproca_new}%%%%where rsproca is the template name

\usepackage{graphicx,color}
\usepackage{dcolumn}% Align table columns on decimal point
\usepackage{url} %this package should fix any errors with URLs in refs.
\usepackage{amsmath,amsfonts,amsthm,bm,amssymb} 
\usepackage{booktabs}
\usepackage{hyperref,comment}
\usepackage{float}
\usepackage{soul}
\usepackage{natbib}
\setcitestyle{numbers,sort&compress}

\begin{document}

%%%% Article title to be placed here
\title{Dynamical Landscape and Multistability of a Climate Model}

\author{%%%% Author details
Georgios Margazoglou$^{1,2}$, Tobias Grafke$^{3}$, Alessandro Laio$^{4}$ and Valerio Lucarini$^{1,2}$}

%%%%%%%%% Insert author address here
\address{$^{1}$Department of Mathematics and Statistics, University of Reading, Reading, United Kingdom\\
$^{2}$Centre for the Mathematics of Planet Earth, University of Reading, Reading, United Kingdom\\
$^{3}$Mathematics Institute, University of Warwick, United Kingdom\\
$^{4}$International School for Advanced Studies (SISSA), Trieste, Italy}

%%%% Subject entries to be placed here %%%%
\subject{Climatology, Statistical Physics, Artificial Intelligence}

%%%% Keyword entries to be placed here %%%%

%%%% Insert corresponding author and its email address}
\corres{Valerio Lucarini\\
\email{v.lucarini@reading.ac.uk}}

%%%% Abstract text to be placed here %%%%%%%%%%%%
\begin{abstract}
We apply two independent data analysis methodologies to locate stable climate states in an intermediate complexity climate model and analyze their interplay. First, drawing from the theory of quasipotentials, and viewing the state space as an energy landscape with valleys and mountain ridges, we infer the relative likelihood of the identified multistable climate states, and investigate the most likely transition trajectories as well as the expected transition times between them. Second, harnessing techniques from data science, specifically manifold learning, we characterize  the data landscape of the simulation output to find climate states and basin boundaries within a fully agnostic and unsupervised framework. Both approaches show remarkable agreement, and reveal, apart from the well known warm and snowball earth states, a third intermediate stable state in one of the two climate models we consider. The combination of our approaches allows to identify how the negative feedback of ocean heat transport and entropy production via the hydrological cycle drastically change the topography of the dynamical landscape of Earth's climate.
\end{abstract}
%%%%%%%%%%%%%%%%%%%%%%%%%%%

%%%%%%%%%% Insert the texts which can accomdate on firstpage in the tag "fmtext" %%%%%

\begin{fmtext}

\end{fmtext}

%%%%%%%%%%%%%%% End of first page %%%%%%%%%%%%%%%%%%%%%

\maketitle

\section{Introduction}\label{sec:Intro}

The climate, an extremely high-dimensional complex system, is determined by five interacting subdomains: a gaseous atmosphere, a hydrosphere (water in liquid form), a lithosphere (upper solid layer), a cryosphere (water in solid form) and a biosphere (ecosystems and living organisms) \cite{Peixoto1992}. The climate is driven by the inhomogeneous absorption of incoming solar radiation and can be treated as a highly non-trivial dynamical system that features spatio-temporal variability on a vast range of scales. The system is at an approximate nonequilibrium steady state  %\textcolor{red}{\st{(except when exhibiting critical transitions)}} 
due to the resulting interplay of forcings, dissipation, positive and negative feedbacks, instabilities and saturation mechanisms \cite{Lucarini2014}. The presence of periodic as well as irregular fluctuations in the boundary conditions does not allow the climate to reach an exact steady state \cite{Ghil2015,Ghil2020}.

A straightforward attempt to mathematically formulate the dynamics of the climate system is by defining a set of partial differential equations (PDEs) that describe the budget of mass, momentum and energy. As this set of PDEs is impossible to solve analytically, they are usually simulated numerically.  Depending on the number of resolved variables, this procedure is extremely challenging both from a technological and scientific point of view, and requires a diversified approach. Therefore, a hierarchy of climate models can be established \cite{schneider1974,Saltzman2001,Held2005,Lucarini2013}. At the lowest level of such a hierarchy one can find simple zero or one-dimensional Energy Budget Models (EBMs) that model the energy exchange in the atmosphere or triggered by the solar radiation \cite{Budyko1969,sellers1969,Ghil1976},	as well as low-dimensional models that represent fundamental processes of the large scale oceanic  \cite{Stommel1961,Veronis1963,Rooth1982} and atmospheric dynamics \cite{Charney1979,Lorenz1984,Lorenz1996}. Next come the so-called intermediate complexity models, which provide a parsimonious yet Earth-like representations of the dynamics of climate, see e.g.~\cite{Marshall1993,fraedrich1998,Petoukhov2000,Montoya2006}. 
Finally, modern state-of-the-art climate models, similar to the ones featured in the latest Intergovernmental Panel on Climate Change (IPCC) report \cite{IPCC2013} are based on applying a series of necessary truncations and approximations in such a set of PDEs. %They simulate, using massive high performance computing resources, a representative amount of physical, chemical, and biological processes of the climate system and attempt to explore its past, present, and future \cite{Randall2000,Ghil2020}.  
In general, the impact of the neglected scales of motions on the explicitly resolved scales is approximated via suitably developed parametrizations, which include deterministic, stochastic, and possibly non-Markovian components \cite{Berner2017}.

\subsection{Global Stability Properties of the Climate System}

The current astronomical configuration of Earth supports the present day Warm (W) climate, and a frozen one, termed Snowball (SB), which exhibits global glaciation, extremely low temperatures and limited climatic variability. Geological and paleomagnetic evidence suggests that during the Neoproterozoic era (in particular around 630 and 715 million years ago), the Earth exhibited at least two major long lasting global glaciation periods, thus entering twice into the snowball climate state \cite{Pierrehumbert2011,Hoffman1998}. Simple energy balance models are able to reproduce the associated multistability of the climate system \cite{Budyko1969,sellers1969,Ghil1976}, which is mainly affected by the so-called ice-albedo feedback. The importance of such a mechanism is confirmed by studies of higher complexity models \cite{Lewis2007,Lucarini2010,Abbot2011,Pierrehumbert2011,Lucarini2017b}.

%Similarly, numerical models of exoplanets suggest the existence of many more stable climates depending on the precise parameters, such as for example 5 co-existing climate regimes on aquaplanets~\cite{Brunetti2019}.

If we now focus on the current climate or the climate of the recent past (thus within the W state), the Earth is well-known to feature further elements of multistability associated with   % which we might define as more local, and is characterized by 
critical transitions among stable states. %Such transitions are often termed
%associated with the so-called tipping points. % in the geophysical community. 
Examples of geographically localized phenomena affecting the climate system featuring such a behaviour -- the so-called tipping elements \cite{Lenton2008} -- include the dieback of the Amazon forest \cite{Boers2017}, the shut-down of the thermohaline circulation of the Atlantic ocean \cite{Rahmstorf2005}, the methane release resulting from the melting of the permafrost \cite{Walter2006}, and the collapse of the atmospheric circulation regime associated to the Indian monsoon \cite{Levermann2009}. %; all of which can be seen as local features of multistability. 
A critical transition taking place for one climatic subsystem may trigger the tipping of another element: this is the phenomenon of so-called cascading tipping points \cite{Steffen2018,Klose2020}. 

Transitions between metastables states might be facilitated by mechanisms like stochastic resonance \cite{Gammaitoni1998}, which has been recently reframed  according to the formalism adopted here for treating nonequilibrium systems \cite{Lucarini2019SR}. Indeed, stochastic resonance  is thought to act in the climate system at different spatial and temporal scales, ranging from ultralong \cite{Benzi1981,Nicolis1982,Ditlevsen2010}, to intermediate \cite{Alley2001,Ganopolski2002,VelezBelchi2001,Lucarini2012}, to short ones \cite{Han2014,ZhengLin2014,Wang2016}.

{\color{black} In this work, we explore the multistability of a climate model through methods from nonequilibrium statistical physics, dynamical systems, and data science, pushing forward the scientific programme presented in \cite{Ghil2020,Lucarini2020} %We aim at providing a viewpoint that is able to describe the multistability of the climate system
%, aiming to broaden the proposed viewpoint on the multistability of the climate system. % and to accommodate the fact that such a multistability appears in a hierarchical fashion. 
 %{\color{black} \st{
 We then take inspiration from the Waddington's ``epigenetic landscape'' metaphor in evolutionary biology~\cite{Waddington1957,Ao2009,Ferrell2012,Huang2012}. }
% {\color{black} In particular, the state space of the climate system is framed within an energy landscape, where 
 {\color{black}The phase space of the climate model can be explored by adding suitably defined stochastic forcing. As a result, the competing metastable climatic states can be viewed as vast valleys of a quasi-potential landscape $\Phi$, separated by mountain ridges, corresponding to unstable climates \cite{Lucarini2019, Lucarini2020}. The stochastic forcing %Random perturbations 
 allows for exploring the landscape and, in particular, makes it possible to observe transitions between the metastable states.} %, which in turn serve as a necessary means of landscape exploration. The estimated relative likelihood within the landscape can be appropriately associated to a global quasi-potential $\Phi$ in the weak noise limit. As it will be demonstrated here, this viewpoint gives a unique perspective to explore the multistable state space of the Earth's climate system through methods from nonequilibrium statistical physics, dynamical systems,  \textcolor{red}{stochastic processes,} and data science.

{\color{black}Unfortunately, the actual evolution of the climate system cannot be fully regarded as the idealised stochastic motion in a fixed nonequilibrium quasi-potential landscape described above because geological, biological, astronomical, and astrophysical factors modulate the landscape on
a vast range of time scales. %, and because the random forcing can be very nontrivial. 
Nonetheless, the quasi-potential landscape viewpoint can be extremely useful to understand its multistability at an instance in time.}

%{\color{black} Finally, we stress that in terms of a realistic climate, unexpected time-dependent geological, biological, astronomical, and astrophysical factors modulate the landscape in ultra-long timescales. Hence, in principle, the actual evolution of the climate system cannot be fully regarded as the idealized stochastic motion in a fixed nonequilibrium quasi-potential,  but instead as a non-stationary landscape that may abruptly change and exhibit nontrivial bifurcations}. \GM{Should we put this here?}
%\st{Unfortunately, the actual evolution of the climate system cannot be fully regarded as the idealised stochastic motion in a fixed nonequilibrium quasi-potential landscape described above because geological, biological, astronomical, and astrophysical factors modulate the landscape on ultralong time scales, and because the random forcing can be very nontrivial.}

	\subsection{Outline of the Paper}% and Main Results}
%Response theory allows to reconstruct locally in the parameters' space the impacts on climate of perturbations to its dynamics \cite{Lucarini2014,Lembo2020}. When applied in the context of time-dependent perturbations of moderate intensity, response theory is key to reconstructing the pullback attractor on which the time-dependent measure of the system is supported \cite{Lucarini2017a,Ghil2020}. % %Nonetheless, the applicability of response theory is inherently limited, as the radius of expansion of this perturbative approach is, by and large, proportional to the spectral gap of the transfer operator of the unperturbed system \cite{Liverani2007,Lucarini2016}. 
%Response theory breaks down, by definition, near critical transitions, where the spectral gap is vanishing and the decay of the correlations becomes very slow \cite{Chekroun2014,Tantet2018}.

In this paper we will study the transitions between competing metastable states of PLASIM \cite{fraedrich2005}, a simplified climate model that has shown extreme flexibility in describing the dynamics of a vast range of climate conditions, including very exotic ones \cite{Lucarini2010b,Boschi2013,Holden2016,ragone2016new,Lucarini2017a,Holden2019}. The model features O($10^5$) degrees of freedom (d.o.f.). %We will see how this allows us to reason about the relative probability of the respective climate states, as well as about the expected times and pathways for transitions between them. 
%\textcolor{red}{\st{Unfortunately, because of the limitations of the model, we will  be able to study multistability only at the largest scale, as indicated with 1 in Fig.}~\ref{fig:multiscale}}. %Indeed, as discussed below, in order to explore the landscape of metastable state, we will force the climate by adding stochastic forcing. However, the intensity of the forcing needed to study effectively the large scale metastability is too strong for retaining the features of the smaller scale ones, which are washed out by noise. In one case we will need to resort to using much smaller noise intensities for studying selectively the decay from a shallow metastable state. This is an example of the beauty and of the curse of multiscale systems: dedicated modelling exercises must usually focus on only one scale of interest, unless one is able to use wisely some form of learning and targeted optimization procedures.

%In order to obtain data to analyse from this novel perspective, we employ %PLASIM \cite{fraedrich2005}, a simplified climate model that has shown extreme flexibility in describing the dynamics of a vast range of climate conditions, including very exotic ones \cite{Lucarini2010b,Boschi2013,Holden2016,ragone2016new,Lucarini2017a,Holden2019}. The model features O($10^5$) degrees of freedom (d.o.f.). 
We consider two setups of the model -- one allowing for the ocean to transport heat from low to high latitudes (setup A), previously used in \cite{Lucarini2017a}, and one where only the atmosphere is able to perform large scale heat transport (setup B), previously used in \cite{ragone2016new}. The main limitation of the model in both setups is its lack of explicit representation of the deep ocean circulation, which is of great relevance for the dynamics of the present climate on centennial to millennial time scales.   

We explore the phase space of the model by allowing the solar irradiance $S^*$ to randomly fluctuate around the present-day mean value of $S^* = 1365 \, \textrm{W}/\textrm{m}^2$, thus triggering transitions among the competing climate states. Along the lines of \cite{Lucarini2019,Lucarini2020}, we  %studying the invariant measure and the statistics of the noise induced transitions among attractors by 
construct the quasipotential of the stochastically perturbed system \cite{Graham1987,Hamm1994,Zhou2012,Zhou2016}. %Such a function, with the caveats discussed below, defines an effective energy landscape for the climate model and is key to understanding the properties of the invariant measure of the system and of the noise-induced transitions. 
and we compute the transitions paths among attractors composed of instantonic and relaxation trajectories by averaging over many transition events.

{\color{black}The identification of the competing attractors is approached in two ways. First, we use standard forward numerical modelling and identify different asymptotic states, associated with separate basins of attraction, when stochastic forcing is removed from the system. Second, competing attractors are automatically detected through data-driven methods applied to the output of long stochastic integrations of the model. Such methods have been used for studying metastable states in biomolecules, and allow one to reconstruct very effectively the quasi-potential $\Phi$ of the system, partially taking care of the \textit{curse of dimensionality} %, and to identify its local minima and saddles 
\cite{faccoID,rodriguez2018computing,automatic-topography,science_DPA}.}
%Following \cite{Tantet2015,Tantet2018}, we construct a finite state Markov chain representation of the transfer operator \cite{Liverani2007,SantosJSP} of the system in a reduced phase space composed of the global surface temperature and the Equator to Poles temperature difference and use it to interpret the intra- and inter-well dynamics.
We anticipate that whereas in setup A we find the two usual W and SB states, we  discover that setup B features the presence of a third stable climate state (to be termed ``cold climate'' (C) in the following), with an ice free latitudinal band at roughly $\pm \, 20^\circ$ around the Equator and mild, larger than $10^\circ$C surface temperatures, along with vigorous atmospheric circulation and non-trivial hydrological cycle in the same band. Such a third state resembles previously suggested exotic climatic configurations such as the slushball Earth \cite{Lewis2007} and the Jormungand state \cite{Abbot2011}. %\textcolor{red}{\st {Note that the interest here is not primarily on the realism of such a state, because of the known limitations on our model, but rather on the fact that its presence makes the statistical mechanics of the problem and the topology of the transition paths both more interesting.}} 
{\color{black}The C state  corresponds to a shallow minimum of the quasipotential and disappears when ocean transport is included in the system, which acts as a strong stabilizing mechanism. The presence of the C state has important implications both on the statistical mechanics of the system and on  the topology of the transition paths between the W and the SB states.}

The paper is structured as follows. Section~\ref{sec:climate_dynamics} contains the mathematical framework behind our analysis. Section~\ref{sec:climate_model} provides a description of the climate model used in this study. Section~\ref{sec:Results} contains the description and critical analysis of the  obtained results. Section~\ref{sec:Conclusions} is dedicated to drawing the conclusion of this work and to presenting future research perspectives. The electronic supplementary material (ESM) attached to this paper, accessible \href{https://doi.org/10.6084/m9.figshare.13079489}{here}, contains some extra information on the computation of the average transition paths and a brief and informal description of the mathematics of the transfer operator and of its finite-size representation. Additionally, it includes a set of movies related to the numerical simulations performed in this study.

\section{Qualitative and Quantitative Aspects of the Multistability of the Climate System}\label{sec:climate_dynamics}
\subsection{Dynamical Landscape of the Climate System}\label{sec:climate_dynamical}

A multidimensional deterministic dynamical system can be defined as a set of ordinary differential equations
\begin{equation}
\frac{\mathrm{d}\bm{x}}{\mathrm{d}t}=\bm{F}(\bm{x},t), \qquad  \bm{x}(t=0)\equiv \bm{x}_0,
\label{eq:detrm}
\end{equation}
where $\bm{x}(t) \in \mathbb{R} ^N$ describes the state of the system at time $t$ with initial condition $\mathbf{x}_0$, and $\bm{F}(\bm{x},t)\in \mathbb{R} ^N$ is a smooth vector field. The initial condition $\bm{x}_0$  determines the asymptotic state of its orbit. If Eq.~\eqref{eq:detrm} possesses more than one asymptotic states, defined by the attractors $\Omega_j$, $j=1,\ldots,J$, the system is multistable. The phase space is partitioned between the basins of attraction $B_j$ of the  attractors $\Omega_j$ and the boundaries $\partial B_l$, $l=1,\ldots,L$ separating such basins, which possess a set of saddle points $\Pi_l$, $l=1,\ldots,L$. Such saddle points attract initial conditions on the basin boundaries \cite{Grebogi1983,Vollmer2009,LT:2011} and can be computed using the so-called edge tracking algorithm \cite{Skufca2006}, which was used in an EBM by \cite{bodai2015}. Chaotic unstable saddles, then termed Melancholia (M) states, have been constructed with the edge tracking algorithm for a simplified climate model built by coupling a primitive equation atmosphere with a diffusive ocean \cite{Lucarini2017b}. 

Escaping an attractor is possible if the system is forced by a properly defined stochastic forcing \cite{hanggi1986,Grassberger1989,freidlin1998} . %The theory of tipping points has been recently extended to deal with all of these scenarios \cite{Ashwin2012}. In particular, noise-induced escapes from attractors have long been studied in natural sciences  \cite{hanggi1986,Grassberger1989} and have, in parallel, inspired extremely fruitful mathematical investigations \cite{freidlin1998}. 
By subjecting Eq.~\eqref{eq:detrm} to a Gaussian random noise and considering it in It\^o form, we write the stochastic differential equation 
\begin{equation}\label{eqapp}
{\mathrm{d}{\bm{x}}}=\bm{F}(\bm{x})\mathrm{d}t+\sigma\bm{s}(\bm{x})\mathrm{d}\bm{W},
\end{equation}
where $d\bm{W}$ is the increment of an $M$-dimensional Wiener process, $\bm{F}(\bm{x})$ is in this context usually referred to as the drift term, $\bm{C}(\bm{x}) = \bm{s}(\bm{x})\bm{s}(\bm{x})^T\in \mathbb{R}^{N\times N}$ is the noise covariance matrix where in general the volatility matrix $\bm{s}(\bm{x}) \in \mathbb{R}^{N\times M}$, and $\sigma\geq 0 $ determines the strength of the noise. 

%The use of stochastic dynamical systems for studying the climate has a long history and has been key to understanding at fundamental level the relationship between climate response to perturbations and climate variability  \cite{Hasselmann1976,Imkeller2001,Saltzman2001,Majda2001,Ghil2015,Ghil2020}.
In the present work, introducing stochasticity in the form of a fluctuating solar constant, amounts to considering  only one independent Brownian motion, so that  $\bm{s}(\bm{x}) \in \mathbb{R}^{N\times1}$ and $C(\bm{x})$ is rank one. Additionally, only the d.o.f.~directly associated to the incoming solar radiation are directly impacted by the stochastic forcing. %We are extremely far from elliptic diffusion processes, where all d.o.f.'s of the system undergo random forcing. 
As clarified in \cite{Lucarini2020}, % and motivated by the way the climate energy exchange processes work, 
we expect that the applied noise percolates to all d.o.f.'s of the system as a result of non-degenerate interplay between stochastic forcing and the deterministic component of the dynamics given by the drift term, so that we can assume that we are dealing with a hypoelliptic diffusion process \cite{Bell2004}. Hence, we expect that for all values of $\sigma>0$ the invariant measure of the system is smooth. 

We now follow \cite{Graham1991,Hamm1994,Zhou2012,Zhou2016}, consider the weak-noise limit, and express the stationary solution of the Fokker-Planck equation corresponding to Eq. \eqref{eqapp} % extend the classic Freidlin-Wentzell's theory \cite{freidlin1998} and allow for studying stochastic processes in nonequilibrium systems featuring multiple deterministic attractors in the form of Eq.~\eqref{eqapp}. In particular, in the limit $\sigma\rightarrow 0$, they 
%introduce the quasipotential $\Phi(\bm{x})$, a nonequilibrium generalization of the notion of free energy, which  characterizes the log-likelihood of reaching a point $\bm{x}$. Indeed, the invariant measure can be written 
as a large deviation law
\begin{equation}\label{eq:stationary_distr}
\rho_\sigma(\bm{x}) \sim Z(\bm{x})\exp\left(-\frac{2 \Phi(\bm{x})}{\sigma^2}\right),
\end{equation}
where $Z(\bm{x})$ is a pre-exponential factor and $\Phi(\bm{x})$ is the quasipotential, a nonequilibrium generalization of the notion of free energy. %, which  characterizes the log-likelihood of reaching a point $\bm{x}$. Indeed, the invariant measure can be written 
$\Phi(\bm{x})$ can be obtained as {\color{black}a nontrivial solution} of the  %By studying a small-$\sigma$ asymptotic expansion of the invariant measure of the system obtained as stationary solution to the Fokker-Planck equation (see Eq.~(1) in Sec.~III of the Supplementary Material and subsequent discussion) Eq.~\eqref{eqa pp}, one obtains 
the following Hamilton-Jacobi equation \cite{Gaspard2002,Zhou2012}:
\begin{equation}\label{eq:HJE}
% W(\mathbf{x}) \sim Z(\mathbf{x})\exp(- \Phi(\mathbf{x})/\sigma^2),
{F}_i(\mathbf{x}) \partial_i \Phi(\mathbf{x})+C_{ij}(\mathbf{x})  \partial_i \Phi(\mathbf{x}) \partial_j \Phi(\mathbf{x}) =0;
\end{equation}
{\color{black}see \cite{Graham1987,Graham1991} for a detailed discussion of the regularity properties of $\Phi$, and \cite{Bouchet2016} for an alternative approach based on variational arguments.} %\textcolor{magenta}{see Sec.~III of the Supplementary Material}. %This equation allows one to express $\Phi$ in terms of the drift and volatility fields introduced in Eq.~\eqref{eqapp}. 
%The quasipotential $\Phi$ can also be computed by solving the variational problem associated with the Freidlin-Wentzell action \cite{Bouchet2016}. We then obtain that %
%The explicit computation of $\Phi$ is far from trivial, yet of great interest in many applications; see e.g., \cite{Zhou2012} for the case of biological systems. Brackston \textit{et al.} \cite{Brackston2018} have recently proposed an algorithm for estimating $\Phi$ in the case that the governing
%equations are polynomial and involves solving an optimization over the coefficients of a polynomial  function. Instead, Tang \textit{et al.} \cite{Tang2017} proposed a variational method for estimating in the  populations corresponding to each deterministic attractor without resorting to computing the invariant measure.
%The function $\Phi$ is key to understanding the properties of the system. Indeed, 
it is possible to write the drift vector field as the sum of two vector fields:
\begin{equation}\label{eq:decomposition}
% W(\mathbf{x}) \sim Z(\mathbf{x})\exp(- \Phi(\mathbf{x})/\sigma^2),
{F}_i(\mathbf{x}) = {R}_i(\mathbf{x})- C_{ij}(\mathbf{x})\partial_j \Phi(\mathbf{x}), \quad {R}_i(\mathbf{x})\partial_i \Phi(\mathbf{x})=0.
\end{equation}
%that are mutually orthogonal, so that  ${R}_i(\mathbf{x})\partial_i \Phi(\mathbf{x})=0$. 
%If Eq.~\eqref{eqapp} describes a thermodynamical system near equilibrium, $\mathbf{R}$ defines the time reversible dynamics, while $\mathbf{F}-\mathbf{R}$ defines the irreversible, dissipative dynamics \cite{Graham1987}. 
A different strategy for attaining the decomposition of the drift term into a symmetric and an antisymmetric component has been proposed by \cite{Ao2004,Yuan2017}.
%
%In general, the decomposition of $\mathbf{F}$ into these two components can be achieved  by solving with respect to $\Phi$ the Note that this equation can also be 
%
%One finds that  
%\begin{equation}\label{eq:Lyap}
%d\Phi(\mathbf{x})/dt = -C_{ij}(\mathbf{x})  \partial_i \Phi(\mathbf{x}) \partial_j \Phi(\mathbf{x}) +  {R}_i(\mathbf{x})  \partial_i  \Phi(\mathbf{x}) = - C_{ij}(\mathbf{x})  \partial_i \Phi(\mathbf{x}) \partial_j \Phi(\mathbf{x}).
%\end{equation}
In the case one switches off the noise, $\Phi$ acts as a Lyapunov function whose decrease with time describes the convergence of an orbit to an attractor. Indeed, $\Phi(\mathbf{x})$ has local minima at the deterministic attractors $\Omega_j$, $j=1,\ldots, J$, and has a saddle behaviour at the saddles  $\Pi_l$, $l=1,\ldots,L$. {If an attractor or saddle is chaotic, $\Phi$ has constant value over its support, which can then be a strange set \cite{Graham1991,Hamm1994}.}   %\GM{With the saddle in the paranthesis you want to say "if an attractor or a saddle is chaotic..."?}

A special class of trajectories, named instantons, define, in the zero-noise limit, the most probable way to exit an attractor   \cite{Kautz1987,Grassberger1989}. An instanton connects an attractor $\Omega$ to a point $\bm{x}$ within the same basin of attraction and can be obtained by minimizing the action of the stochastic field theory associated with the system \cite{Grafke2015,Bouchet2016,Grafke2017,Grafke2019}. The instantonic trajectory obeys the equation of motion $\mathrm{d}x_i/\mathrm{d}t = R_i(\bm{x}) + C_{ij}(\bm{x}) \partial_j \Phi_\Omega(\bm{x})$, which has a reversed component of the gradient contribution with respect to the drift field, see Eq.~\eqref{eq:decomposition}.  If $\bm{R}(\bm{x})$ vanishes, instantonic trajectories follow the same path (in reverse direction) with respect to relaxation trajectories, which is a basic characterization of equilibrium systems and detailed balance. %The instanton is intimately connected to the quasipotential $\Phi(\bm{x})$ 

Within the basin of attraction of $\Omega$ one can define the local quasipotential $\Phi_\Omega(\bm{x})$ as the action for the instanton linking $\Omega$ and $\bm{x}$ \cite{Bouchet2016}. %The instantonic trajectory obeys the equation of motion $\mathrm{d}x_i/\mathrm{d}t = R_i(\bm{x}) + C_{ij}(\bm{x}) \partial_j \Phi_\Omega(\bm{x})$, which has a reversed component of the gradient contribution with respect to the drift field, see Eq.~\eqref{eq:decomposition}. If $\bm{R}(\bm{x})$ vanishes, instantonic trajectories follow the same path (in reverse direction) with respect to relaxation trajectories, which is a basic characterization of equilibrium systems. 
%To recover the global quasipotential $\Phi(\bm{x})$, one needs to resort to a pruning-and-stitching strategy, glueing together the local portions $\Phi_{\Omega_j}, j=1,\ldots,J$, see \cite{Graham1991} and the careful description recently provided by~\cite{Zhou2016}. A separate view on this problem, based upon a different interpretation of the noise has been proposed in \cite{Ao2004,Yuan2017}.
%Instantonic trajectories allow to reason about expected escape times in the zero-noise limit. 
Escapes from an attractor $\Omega$ take place through a saddle $\Pi$ situated at the boundary of the basin of attraction having the lowest value of the local quasipotential barrier height $\Delta \Phi_{\Omega\to \Pi}=\Phi_\Omega(\Pi)-\Phi_\Omega(\Omega)$~\cite{LT:2011} and are Poisson-distributed events, where the probability that an orbit does not transition up to time $t$ is, similarly to the classic Kramers' law \cite{Kramers1940}, given by: 
\begin{equation}\label{eq:tt_distr}
P(t) =\frac{1}{\bar{\tau}_\sigma}\exp\left(-\frac{t}{\bar{\tau}_\sigma}\right), \, \mathrm{with}\, \bar{\tau}_\sigma \propto \exp\left(\frac{2\Delta \Phi_{\Omega\to \Pi}}{\sigma^2}\right),
\end{equation}
%being the expected escape time and $\Delta \Phi_{\Omega\to \Pi}=\Phi_\Omega(\Pi)-\Phi_\Omega(\Omega)$ is the quasipotential barrier height at the relevant saddle~\cite{LT:2011}. 
Unfortunately, in the case of multistable systems, one cannot, in general, simply read off the barrier height $\Delta \Phi_{\Omega\to \Pi}$ from the $\Phi(\bm{x})$ of Eq.~\eqref{eq:stationary_distr}, because glueing together the various local quasipotentials does not give the global quasipotential $\Phi(\bm{x})$  \cite{Graham1991,Zhou2016}. %While $\Phi(\bm{x})$ yields information about the relative probability of attractors, and is available e.g.~through global sampling of the system, the local notion of potential barriers, $\Delta\Phi_{\Omega\to \Pi}(\bm{x})$ is relevant for the time-scale of transition events, and can be obtained e.g.~by looking at transition times between attractors.
The local and global notions of quasipotential can be brought to a common ground if the system is at equilibrium so that no global probability fluxes are present. Equivalence between the information provided by the local and global quasipotentials is also realized  if the system is not an equilibrium one but only two competing states are present {\color{black}with a single saddle embedded in the boundary between the two basins of attraction}, as in the cases analysed in \cite{Lucarini2019,Lucarini2020}. In general, we will resort to measuring separately the invariant measure~\eqref{eq:stationary_distr} and the barrier heights~\eqref{eq:tt_distr}.

%	however, if these sets are not fixed points, the instanton is not unique, as the quasipotential is constant on $\Omega_j$ and $\Pi_l$.  Instantons are defined as solutions of 
%\begin{equation}\label{eqappi}
%d{{{x}}_i}/dt={\tilde{F}}_i(\mathbf{x})={R}_i(\mathbf{x})+C_{ij}(\mathbf{x})\partial_j \Phi(\mathbf{x})
%\end{equation}
%that connect a point in $\Omega_j$ to a point in $\Pi_l$. Instantonic trajectories have a reversed component of the gradient contribution to the vector field compared to regular trajectories.

\subsection{Exploring the topography of the quasipotential}\label{sec:topography}
%A key role in our analysis is played by the quasipotential $\Phi$ defined in Eq.~\eqref{eq:stationary_distr}, {\color{black}because, as mentioned above, the topography of  $\Phi$ determines the macroscopic features of the dynamics.} % The  number of its  minima is equal to the number of metastable states, and the height of the barriers between these minima (approximately) determines the time scales of the transitions. 
%In general a multiscale landscape will be characterized by a relatively few large basins, separated by high barriers, decorated by smaller local minima like in the example of Fig.~\ref{fig:multiscale} . 

To  study the topography of $\Phi$ one can neglect the preexponential factor $Z(x)$ in Eq.~\eqref{eq:stationary_distr} and  project the invariant measure $\rho_{\sigma}(x)$ on a small number $n$ of pre-selected variables defined by the function $s=S(x)\in \mathbb{R}^n$. This gives 
\begin{equation}
\Phi(s) \sim -\frac{\sigma^2}{2} \log \rho_{\sigma}(s)= -\frac{\sigma^2}{2} \log \int dx \delta(S(x)-s) \rho_{\sigma}(x).
\label{eq:dimred}
\end{equation}
If $n$ is small, $\rho_{\sigma}(s)$ can be efficiently estimated, e.g., by computing a histogram. %, {\color{black}see, e.g., Figs.~\ref{fig:ocdiff_1_phsp} and~\ref{fig:ocdiff_0_phsp}.} 
Its minima and saddle points can then be found straightforwardly, even by visual inspection. However, this approach has an important drawback: the choice of the variables on which one projects is arbitrary, and multiple attractors may appear erroneously merged into a single one for a too low-dimensional projection, as shown later. % We will show later -- see Fig.~\ref{fig:ocdiff_0_phsp} -- that the two-dimensional quasipotential of a realistic climate model estimated as a function of two \emph{reasonable} variables has two clear local minima in a case in which the attractors are three.   

To circumvent this problem one can perform the analysis with an approach borrowed from manifold learning, which allows estimating the quasipotential as a simultaneous function of a large number of variables % ({\color{black}we will limit ourselves to} $\mathcal{O}(100)$ in the case of this work, {\color{black} but the approach can be used seamlessly for many more}) 
and studying its topography directly in this space. As we will show, this allows one to identify the deterministic attractors of a system of the form given in Eq. \eqref{eqapp} without preselecting a small number of putative important variables, i.e.~it is applicable even when $n\gg 1$.

This procedure is rooted on a pretty general property of dynamic{\color{black}al} systems: even if the dynamics takes place in a $D$-dimensional space, where $D$ can be very large, the trajectory is often contained in an embedding  manifold of dimension $d$ where  typically $d\ll D$ \cite{fractal_dim}; {\color{black}in the case of deterministic chaos this information in encoded by the Kaplan-Yorke dimension \cite{Frederickson1983}.}  
This, as we will see,  makes  the estimate of $\rho_{\sigma}$ \emph{restricted to the manifold} numerically and algorithmically possible.  However, this manifold is typically twisted and curved, and it is very difficult (or even impossible, if the topology of the manifold is non-trivial) to define a global coordinate chart. %In other words, a vector function $s=S(x)$ which is appropriate for estimating the quasipotential from  Eq.~\eqref{eq:dimred} typically exists (and  is $d$-dimensional) but it is very difficult finding it in practice.  
The approach we use allows one to estimate the quasipotential directly on the embedding manifold as in Eq. \eqref{eq:dimred} without defining explicitly the function $S(x)$.

Consider a trajectory $\mathbf{x}_t$, where $t$ labels the different configurations.  Consider the Euclidean distance  $r_{t,t'}=\Vert \mathbf{x}_t-\mathbf{x}_{t'}\Vert$ between pairs of configurations. Even if this distance is defined in a $D$-dimensional space, if  $\mathbf{x}_t$ and $\mathbf{x}_{t'}$ are so close that one can neglect the curvature, $r_{t,t'}$ approximates a metric on the manifold. 
Building on this approximation, one first estimates  $d$  from the statistics of the ratio between the distance of the closest neighbor $r_{t,(1)}$ of each data point $t$ and the distance of its second nearest neighbour $r_{t,(2)}$. 
One can prove that $\mu_t=\frac{r_{t,(2)}}{r_{t,(1)}}$ is Pareto distributed \cite{faccoID}: $\mu_t \sim \mathrm{PD}(d)$, except for a correction which depends on  the curvature of the manifold and on the variation of the invariant measure on the scale of distance $r_{t,(2)}$. These errors vanish in the limit of infinite sample size \cite{faccoID}.
This allows inferring the value of $d$ from the empirical probability distribution of $\mu$; see closely related results in  \cite{Lucarini2016Extremes,Faranda2017}.

The next step is estimating %$\rho_{\sigma}$ in each configuration $x_t$, or equivalently, 
the quasipotential $\Phi_t=-\frac{\sigma^2}{2} \log(\rho_{\sigma}(x_t))$. This is done using the approach in Ref. \cite{rodriguez2018computing}, a generalization of the $k$-Nearest Neighbor density estimator \cite{knn} in which the probability density is estimated implicitly on the embedding manifold and the optimal $k$ becomes configuration-dependent. The optimal $k$  is defined by finding, via a statistical test, the largest neighborhood of $x_t$ in which the density can be considered constant with a given statistical confidence. We denote by $\mathcal{N}_t$ this neighborhood and by $\hat{k}_t$ the optimal value of $k$ for configuration $t$. $\Phi_t$ is then obtained by maximizing a likelihood with respect to two variational parameters\cite{rodriguez2018computing}:
\begin{equation}
\Phi_t = \underset{\phi}{\operatorname{argmax}} \; \max_{a} \left( 
-\phi \hat{k}_t+a\frac{\hat{k}_t(\hat{k}_t+1)}{2}
-\sum_{l=1}^{\hat{k}_t} e^{-\phi+al} v_{t,l}
\right)
\label{eq:F_PAk}
\end{equation}
where, denoting by $\Omega_d$  the volume of a $d$-sphere of unitary radius and by $r_{t,(l)}$ the distance between $x_t$ and its $l$-th nearest neighbour, $v_{t,l}=\Omega_d \left( r_{t,(l)} ^d- r_{t,(l-1)}^d\right)$. 
%This procedure requires estimating in advance the value of $d$, since this parameter enters in the expression of the hyperspherical shells $v_{t,l}$.  
Importantly, this procedure provides, within the same statistical framework used for defining the likelihood in Eq.~\eqref{eq:F_PAk}, an estimate of the error on $\Phi_t$, which we denote by $\varepsilon_t$. %As we will see, this allows assessing quantitatively the statistical reliability of the partition in attractors derived by the model. 

The final step is inferring the topography of the quasipotential from the estimates $\Phi_t$. This is done through an unsupervised extension of Density Peak Clustering \cite{automatic-topography,science_DPA}. 
Configuration  $t$ is assumed to be a local minimum of $\Phi$ if the following two properties hold: (I) $\Phi_t<\Phi_{t'} \; \forall x_{t'} \in \mathcal{N}_t$, namely if $x_t$ is a minimum in $\mathcal{N}_t$,  (II) $x_t\notin \mathcal{N}_{t'} \; \forall t' : \Phi_{t'}<\Phi_t $, namely if  $x_t$ does not belong to the  $\mathcal{N}_{t'}$ neighborhood of any configuration with lower $\Phi$.
An  integer label $c$ is assigned to each of the $n$ local minima found in this manner. The labels of the other configurations are found iteratively, by assigning to each point the same label of its nearest neighbor of smaller $\Phi$ \cite{science_DPA}.

The set of points with the same label $c$ is denoted by $\mathcal{A}_c$ and is assumed to correspond to a basin of attraction.
The saddle points between the attractors  are then found.
A configuration  $ x_t \in \mathcal{A}_c$ is assumed to belong to the border  with a different attractor $\mathcal{A}_{c'}$ if there exists   a configuration $x_{t'} \in \mathcal{N}_t \cap \mathcal{A}_{c'}$ such that $r_{t,t'}=\min_{x_{t"} \in \mathcal{A}_{c}} r_{t",t'}$.
The saddle point between $\mathcal{A}_{c}$ and $\mathcal{A}_{c'}$ is the point   of minimum $\Phi$ belonging to the border between the two attractors.

Finally, the statistical reliability of the attractors is assessed  as follows.
Denote by $\Phi_c$ the minimum value of $\Phi$ in the attractor $c$, by $\varepsilon_c$ its error, by  $\Phi_{c,c'} $ the value of $\Phi$ of the saddle point between $\mathcal{A}_{c}$ and $\mathcal{A}_{c'}$ and by $\varepsilon_{c,c'}$ its error.
If $\Phi_{c,c'}-\Phi_{c'} < Z \sqrt{\varepsilon_c^2+\varepsilon_{c,c'}^2} $, the attractor $c'$ is merged with attractor $c$ since the value of the quasipotential at its minimum  and at the saddle point are indistinguishable at a statistical confidence defined by $Z$ \cite{automatic-topography}.
The process is repeated until all the attractors satisfy this criterion, and are therefore statistically robust with a confidence $Z$. 

{\color{black} The whole procedure enables us to detect metastable states that might be masked in any low dimensional projection of the invariant measure. In the case the analysed data have been produced using a numerical model (as is the case here), it is possible to have conclusive results on the correctness of a candidate attractor by running noiseless forward simulations from the best estimate of its position (and nearby points) and observe whether it indeed persists indefinitely.}

\section{The climate model}\label{sec:climate_model}

We perform the numerical simulations using PLASIM, an open-source intermediate complexity climate model developed at the University of Hamburg \cite{fraedrich2005}. PLASIM has a total of $\mathcal{O}(10^5)$ d.o.f., and retains some of the most important features of the climate, but is considerably less sophisticated and cheaper to run than the present state-of-the-art Earth System Models that reach more than $\mathcal{O}(10^8)$ d.o.f. \cite{Eyring2016}. PLASIM is extremely flexible and has been used for studying a rather wide range of climatic conditions \cite{Lucarini2010b,Boschi2013,Holden2016,ragone2016new,Lucarini2017a,Holden2019}, hence providing the perfect testing ground for novel theoretical investigations in climate science. PLASIM is well-known to feature multistable dynamics, which has been thoroughly studied in previous studies \cite{Lucarini2010,Lucarini2013,Boschi2013}. 

The dynamical core of PLASIM is responsible for describing the mass and the budgets of momentum, energy, and water in the atmosphere. % and is based on its predecessor model, PUMA \cite{fraedrich1998}.
The primitive equations are solved by the spectral transform method \cite{Orszag1970} in the horizontal, by finite differences in the vertical and for the time advancing scheme, a semi-implicit time stepping is used \cite{haltiner1980}. Further to that, unresolved physical processes, e.g.~horizontal and vertical diffusion, long and short wave radiation, interaction with clouds, moist processes and dry convection, precipitation, boundary layer fluxes of latent and sensible heat, and a land surface with biosphere are among the many to be effectively parameterized into the model. In that way, PLASIM  simulates with a fair degree of accuracy all the necessary components of a realistic Earth-like climate system, with the notable exception of a dynamical component able to simulate the deep oceanic circulation; see discussion in \cite{Lembo2020}. As it will become apparent below, the presence in PLASIM of a reasonably realistic representation of the hydrological cycle is key to introducing a new layer of complexity in the present study compared to what had been explored in previous investigations of the global stability properties of the climate systems \cite{Lucarini2017b,Lucarini2019,Lucarini2020}. %Compared to such studies, the model used here is much closer to an actual state-of-the-art climate model as it includes -- yet in a somewhat simplified way --  a much more sophisticated representation of processes like radiation, hydrological cycle, clouds, water phase changes, and surface processes.   %, by emphasizing the nonequilibrium properties of the system we are investigating.

%\subsection{Experimental configuration}\label{sec:Experimental configuration}

Our experimental configuration uses a present day geography and further consists of an oceanic mixed layer of 50~m depth via a one-layer slab ocean model, which includes a thermodynamic sea-ice module \cite{Semtner1976}. %Furthermore, we neglect contributions from continental ice-sheets, vegetation and interactive carbon cycle. 
The resolution of the model is T21 in the horizontal direction, corresponding to a $5.6^\circ \times 5.6^\circ$ grid cell, with 10 levels in the vertical, while the time-step is 45~min. Finally, we fix the CO$_2$ concentration to 360 ppm, while daily and seasonal cycles have been purposefully neglected to further remove any explicit time dependency of the evolution equations.

We configure two experimental setups that differ in terms of how the oceanic heat transport is prescribed. In setup A the horizontal ocean diffusion is active and its parametrization requires choosing a specific value for the horizontal diffusivity constant. This setup allows for a simple yet effective representation of the impact of the large-scale ocean transport on the climate as a whole,  and has been used in a recent study where response theory was used to perform climate projections \cite{Lucarini2017a}. In setup B, the horizontal ocean diffusivity is set to 0, which implies that the associated feedback to the large scale heat transport performed by the ocean is neglected. A similar configuration as in setup B has been previously employed to study the thermodynamic properties of the climate system in response to controlled changes of the solar constant \cite{Lucarini2010} or of the CO$_2$ concentration \cite{Lucarini2010b,ragone2016new}.

%In order to explore the phase space of the system and its competing metastable states, we consider, as in \cite{Lucarini2019,Lucarini2020}, random fluctuations of the solar irradiance $S^*$. % in a similar spirit as in . %Compared to such studies, the model used here is much closer to an actual state-of-the-art climate model as it includes -- yet in a somewhat simplified way --  a much more sophisticated representation of processes like radiation, hydrological cycle, clouds, water phase changes, and surface processes. %An -- important -- missing element in PLASIM is the representation of slow scale processes associated to the dynamics of the deep ocean; see discussion in \cite{Lembo2020}.
Following \cite{Lucarini2019,Lucarini2020}, the stochastic forcing needed to explore the phase space of the system is introduced here as random fluctuations of the solar irradiance around its present value $S^*_0 = 1365 \, \textrm{W}/\textrm{m}^2$. Each year a different value is prescribed according to $S^*=S^*_0 +\eta$ , where $\eta$ is a random number drawn from a normal distribution with vanishing mean and standard deviation $\delta S=\sigma S^*_0$. We consider a vast range of values for $\sigma$, ranging from $0.01$ to $0.26$, and perform multiple simulations ranging from hundreds to tens of thousands of years, in order to explore at different level of accuracy the local as well as the global properties of the phase space of the system.  %different standard deviation $\delta S$, defined as a fraction of the mean $\delta S = \sigma \, S^*$, where $0 < \sigma < 1$. 
%Spanning from $5\cdot10^{3}$ to $6\cdot10^{4}$ years duration, this recipe allows to sufficiently explore the phase space of the system for different noise intensities. 
Note that when weaker noise is considered, the exploration of the phase space requires longer integrations, as the transitions between the basins of attraction become more rare, see discussion below.

\section{Results}\label{sec:Results}

\subsection{Setup A -- Atmospheric and Oceanic Large Scale Energy Transport} \label{sec:Physics and climatology of setup B}

\subsubsection{The Two Competing Climate States}

In setup A %of the model is such that the slab ocean is able to transport energy poleward thanks to the presence of a horizontal diffusion. 
the representation of the large scale oceanic energy transport is, euphemistically, oversimplified compared to what really occurs in Earth, as our model cannot represent the process of deep water formation and the large scale circulation of the ocean \cite{Dijkstra2005,Kuhlbrodt2007,Cessi2019}. Nonetheless, the presence of horizontal heat diffusion performed by the ocean has the great merit of introducing an additional mechanism -- on top of atmospheric transport fuelled by baroclinic instability -- that contributes to reducing the large scale temperature difference between low and high latitudes \cite{Peixoto1992,Trenberth2009,Lucarini2011,Knietzsch2015}. We find, as expected, two competing asymptotic states corresponding to the W and SB climates, in agreement with a plethora of previous investigations, as discussed in the introduction.

\begin{figure}[b]
	\centering		
	\includegraphics[width=0.24\linewidth]{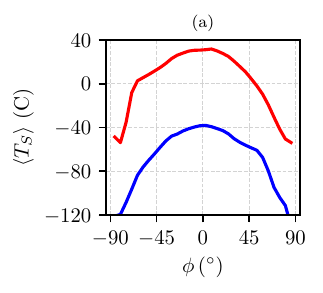}
	\includegraphics[width=0.24\linewidth]{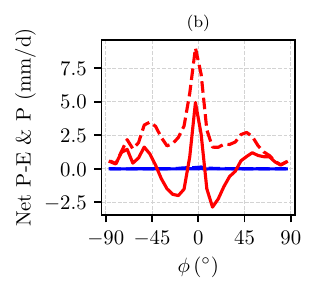}
	\includegraphics[width=0.24\linewidth]{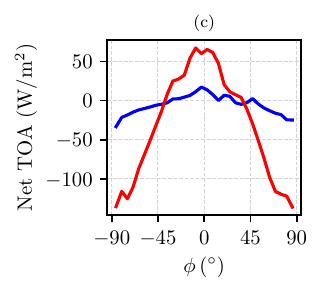}
	\includegraphics[width=0.22\linewidth]{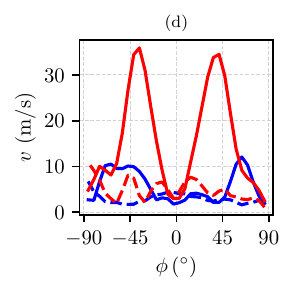}
	\caption{Climatological longitudinally averaged annual mean of (a) surface temperature, (b)  Precipitation minus Evaporation (P-E, solid lines) and Precipitation (P, dashed lines), (c) top of the atmosphere net radiation,  (d) magnitude of zonal wind speed at 300 hPa (solid lines) at 1000 hPa (near surface, dashed lines) versus the latitude $\phi$. Blue lines: SB state. Red lines: W state.}
	\label{fig:climatology_setupA} 
\end{figure}

In Fig.~\ref{fig:climatology_setupA} we present the zonally averaged annual mean of a 40-year long time-series of several observables, computed when steady state conditions are realized in absence of stochastic forcing  ($\sigma=0$). We compare here zonally averaged fields of the W climate (red lines) and of the SB (blue lines); additional information on globally averaged quantities are presented in Table~\ref{tab:characteristics}.

\begin{figure}[ht]
	\centering
	\includegraphics[width=0.8\linewidth]{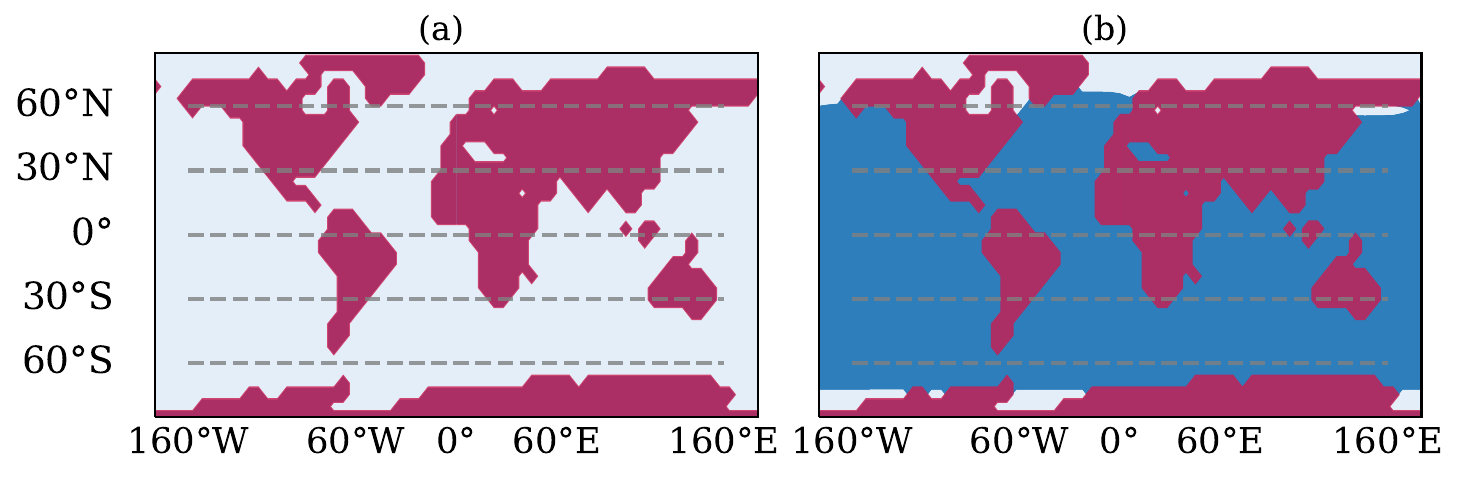}
	\caption{Sea-ice coverage comparison between (a) snowball and (b) warm climates, where the color coding is white for ice, blue for sea and red for land. We depict the land-map used by our model.}
	\label{fig:maps_setupA} 
\end{figure}

Fig.~\ref{fig:climatology_setupA}(a) shows the climatology of the zonal mean surface temperature. In agreement with previous studies performed on PLASIM \cite{Lucarini2010,Lucarini2013,Boschi2013}, the SB state features global glaciation and extremely low temperatures at all latitudes, while the W state is similar to the present-day climate; see also the map of sea-ice cover in Fig.~\ref{fig:maps_setupA}, where the limit of sea-ice approximately coincides with the isoline of $0^\circ$ C in the surface temperature shown in Fig.~\ref{fig:climatology_setupA}(a).   %albeit considerably colder; see more on this point in the later discussion of the climatology of setup B. %This can be further understood by the global distribution of the sea-ice coverage  Clearly, the C state exhibits an ice-free equatorial belt, compared to the fully ice-covered SB state, or the W state, where sea-ice reaches latitudes of $\phi=\pm50^\circ$. 

Figure~\ref{fig:climatology_setupA}(b) shows the annual mean budget of the precipitation minus evaporation rate (P-E) as well as the annual zonally averaged precipitation. The SB climate is almost entirely dry, as a result of the fact that the very low temperature of the atmosphere permits the presence of nothing but an extremely small amount of water vapour, because of the constraint posed by the Clausius-Clapeyron relation \cite{Peixoto1992}. The W climate has the familiar maximum of precipitation in the equatorial belt and secondary peaks in the mid-latitudes, resulting from convective precipitation and synoptic disturbances, respectively. The  P-E field describes the scenario of net water vapour transport from the tropics into the equatorial belt and into the mid-latitudes \cite{Peixoto1992}. %The C state features an intense precipitation in the equatorial belt, driven by the strong convection occurring there, but the net EP indicates that the water vapour is recycled and no large scale transport takes place. 

Figure~\ref{fig:climatology_setupA}(c) shows the zonally averaged net TOA energy budget, which is  the sum of the incoming shortwave radiation and the outcoming longwave radiation and scattered shortwave radiation. Note that the fluxes are positive when entering the planet and negative when leaving the planet. At steady state, the zonal TOA energy imbalance is compensated by the divergence of the  meridional atmospheric enthalpy transport \cite{Peixoto1992,Trenberth2009,Lucarini2011}. We then conclude  that such a transport is much stronger for the W state, where large contributions come from baroclinic eddies and from the large scale transport of water vapour. Baroclinic eddies are located in the region of the jet, where zonal winds in the upper troposphere at 300 hPa (near the tropopause, where the peak intensity is found) -- Fig.~\ref{fig:climatology_setupA}(d) -- and their existence is made possible by the conversion of available potential into kinetic energy via baroclinic instability, which is associated to the presence of a substantial meridional temperature difference between low and high latitudes in the atmosphere. The vigorous circulation of the W state corresponds to a  powerful Lorenz energy cycle \cite{lorenz1967} ($\approx 3.4 \text{W m}^{-2}$). %v equally intense in the W climate %and C climates, yet the peaks of the transpoery intense in rt -- indicated by vanishing values of the TOA budget \cite{Peixoto1992,Trenberth2009,Lucarini2011}- are confined to lower latitudes in the latter case. This indicates a vigorous heating realized at $\approx \pm 30^oN$. As a result, the position of the jet stream is at lower latitudes compared to the W climate (panel d), yet with similar magnitude, reflecting the presence of comparable meridional temperature gradients in the atmosphere. 
Instead, the meridional enthalpy transport and the zonal circulation of the SB state are extremely weak, corresponding to the presence of very modest meridional temperature gradients \cite{Lucarini2010,Lucarini2013,Boschi2013}. The SB state features a very weak Lorenz energy cycle ($\approx 1.0 \text{W m}^{-2}$), as the presence of a weak meridional temperature gradient leads to a scarce reservoir of available potential energy and shuts down almost entirely the mechanism of baroclinic instability. The vast difference in the intensity of the Lorenz energy cycle in the two climates corresponds to the presence of much weaker surface winds in the SB than in W climate; see Fig.~\ref{fig:climatology_setupA}(d).

\begin{table}[ht!]
	\caption{Main climatic features of the stable climates for the two experimental configurations \textcolor{black}{in absence of stochastic forcing ($\sigma=0$)}, where A refers to setup A, and B to setup B; W for warm state, C for cold state, and SB for snowball state; LEC stands for Lorenz energy cycle.}
	\centering
	\def\arraystretch{1.2}
		\begin{tabular}{l|cccc}
			\hline
			& $[\langle T_S\rangle]$ ($^\circ$C) &$\Delta T_{EP}$ ($^\circ$C) & sea ice (\%) & LEC (W/m$^2$) \\ \hline
			A W  & 15.0(2)  & 26.4(3)  & 5.5(1)  & 3.39    \\
			A SB & -55.2(3)  & 25.7(5) & 100   & 1.00  \\ 
			B W  & 4.4(3)  & 40.0(5) &  27.7(1)   & 4.79    \\
			B C  & -28(2)   & 53(1)  & 70(2)   & 3.79    \\
			B SB & -52.5(5)  & 25.9(5) & 100   & 1.19  \\ \hline
		\end{tabular}
	\label{tab:characteristics}
\end{table}

\subsubsection{Noise-induced Transitions}

In what follows, we will apply a very severe coarse-graining to the phase space of the model. Indeed, we perform a projection on the plane spanned by the  globally and 30-day averaged surface temperature $[\langle T_S\rangle]$ and 30-day averaged Equator minus Poles surface temperature difference $\Delta T_{EP} = \langle T_{Eq}\rangle - \langle T_{Po}\rangle$, where we denote the spatial average of the field X by $[X]$, and the temporal average by $\langle X\rangle$. Specifically,  $T_{Eq} = [ T_S]_{30^\circ S}^{30^\circ N}$ and  $T_{Po} \equiv \left([ T_S]_{30^\circ N}^{90^\circ S}+[T_S]_{90^\circ S}^{30^\circ S}\right)/2$. Such a projection allows retaining a minimal yet still physically relevant description of the system \cite{bodai2015, Lucarini2017b,Lucarini2019,Lucarini2020}. Indeed, variations in the globally averaged surface temperature reflect, to a first approximation, changes in the energy budget of the planet (warming vs cooling), while $\Delta T_{EP}$ controls the large scale energy transport performed by the geophysical fluids \cite{Peixoto1992,Lucarini2011}. 

\begin{figure}[]
	\centering
	\includegraphics[width=.49\linewidth]{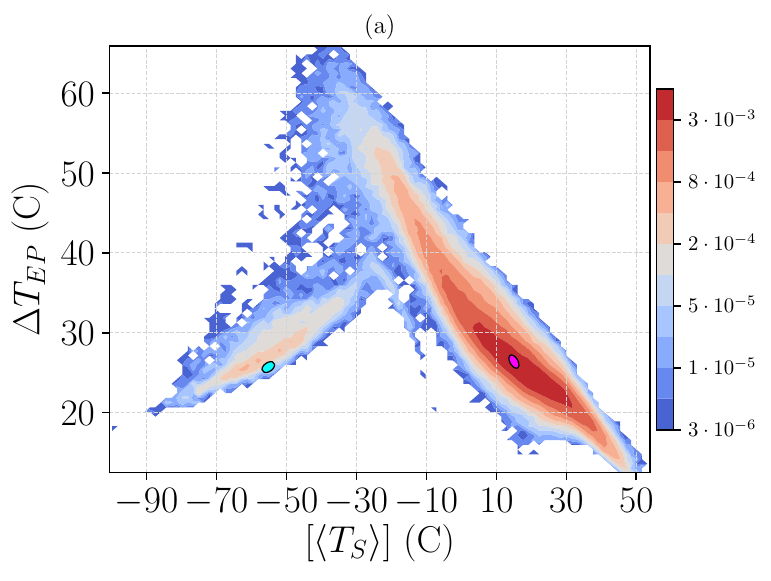}
	\includegraphics[width=.49\linewidth]{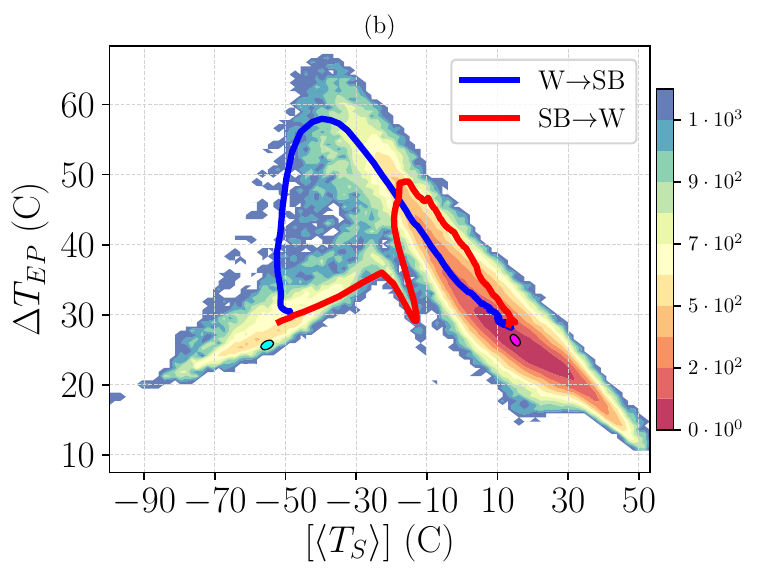} 
	\includegraphics[width=.49\linewidth]{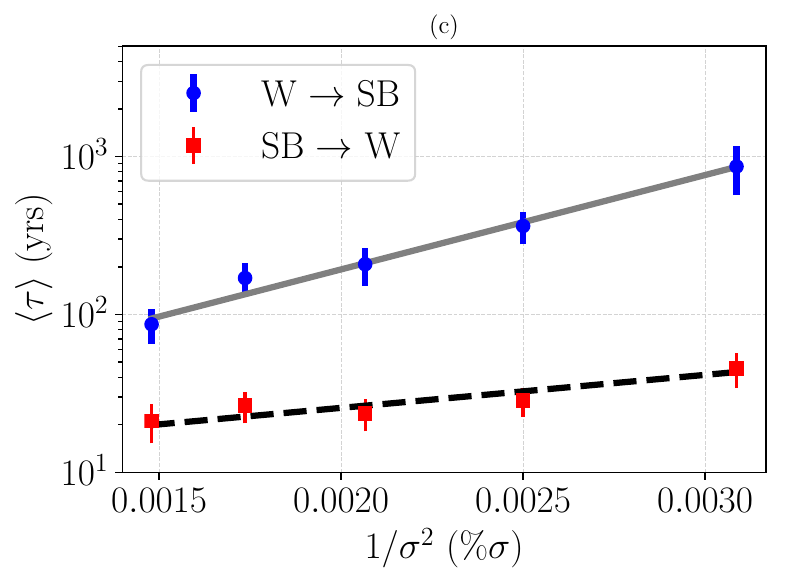} 
	\includegraphics[width=0.5\linewidth]{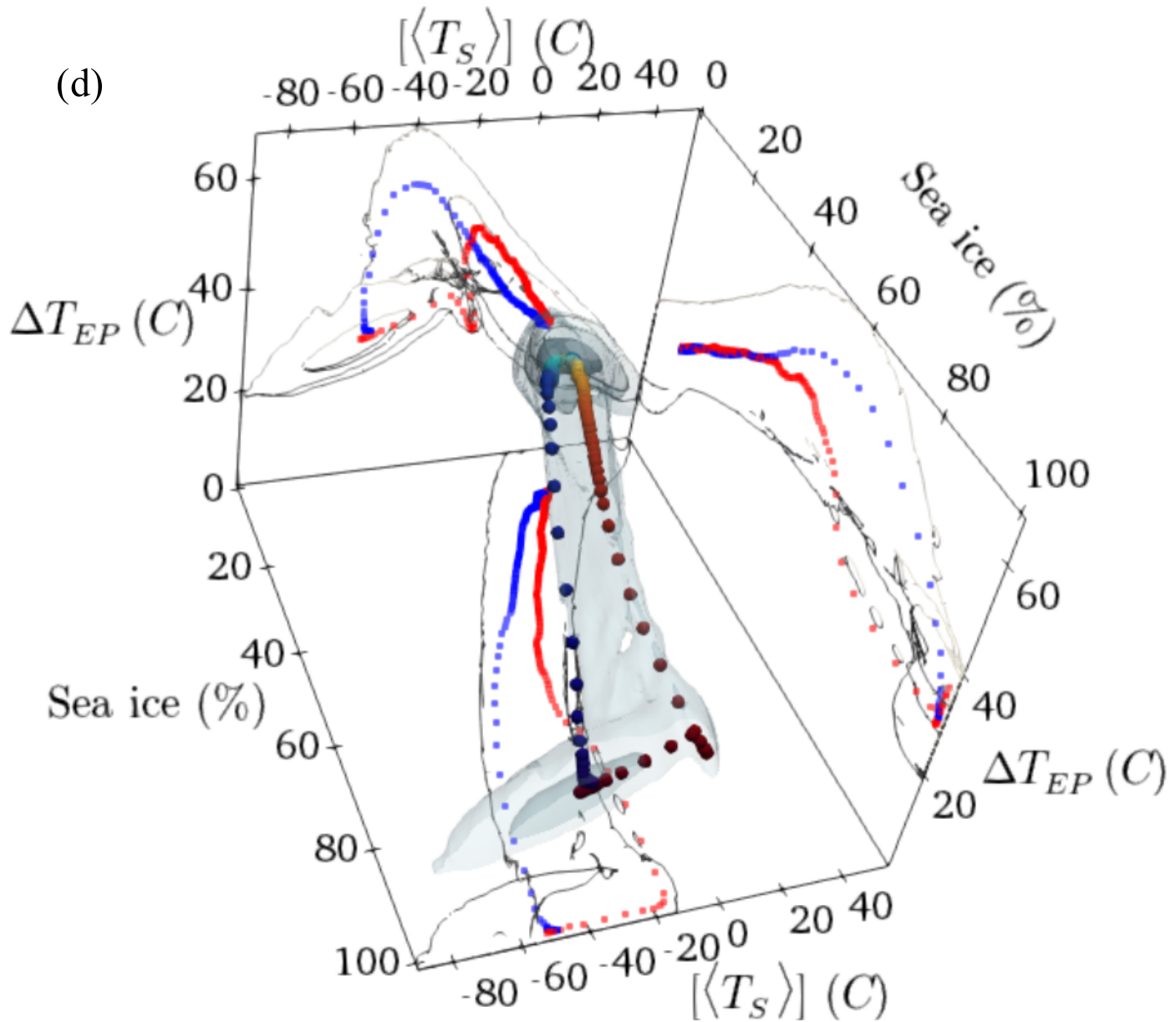}	
	\caption{ (a) Projection of the invariant measure on the reduced phase space spanned by the 30-day averaged global temperature $[\langle T_S\rangle]$  versus 30-day averaged Equator to pole temperature difference  $\Delta T_{EP}$ for $\sigma=18\%$. (b) The corresponding quasipotential, shifted so that the global minimum is set to 0. The blue and red lines correspond to stochastically averaged transition paths  for the W$\rightarrow$SB and SB$\rightarrow$W transitions. These trajectories follow different escape and relaxation paths, within each basin of attraction. The colored ellipses indicate the location of the deterministic attractors corresponding to SB state (cyan), and W state (magenta). 
	(c) Average escape time versus the inverse squared $\%\sigma$, where dashed black and straight gray lines correspond to fitting Eq.~\eqref{eq:tt_distr}. (d) Transition paths $\rm SB\rightarrow W$ (red) and $\rm W\rightarrow SB$ (blue) in the 3D space spanned by $[\langle T_S\rangle]$, $\Delta T_{EP}$, and the arctic sea ice percentage for $\sigma=18\%$. The shading indicates the density of the projected invariant measure, while a 2D projection of the transition paths in each plane is added. All data refer to setup A.}% and noise intensity $\sigma=18\%$. }
	\label{fig:ocdiff_1_phsp} 
\end{figure}

The asymptotic state of the system in absence of any form of stochastic forcing corresponds to either of the attractors described above and is determined by the initial condition. Transitions between the attractors can be induced by noise. 
In Fig.~\ref{fig:ocdiff_1_phsp}(a) we present the projection of the invariant measure of the stochastically forced system ($\sigma = 18 \%$) on the reduced phase space spanned by $[\langle T_S\rangle]$ and $\Delta T_{EP}$ (normalized to one), while Fig.~\ref{fig:ocdiff_1_phsp}(b) portrays the % following  \cite{Gaspard2002,Zhou2016}, we approximately estimate the corresponding 
quasipotential estimated using Eqs.~\eqref{eq:stationary_distr} and \eqref{eq:dimred}:
\begin{equation}\label{eq:quasipotential}
\Phi([\langle T_S\rangle],\Delta T_{EP}) \sim - \frac{\sigma^2}{2}\log \rho_\sigma([\langle T_S\rangle],\Delta T_{EP}),
\end{equation}
where the global minimum is set to 0. %from $\Phi([\langle T_S\rangle],\Delta T_{EP})$ to shift to zero the  level of the deepest quasi-potential well. 
Note that the noise level given by $\sigma = 18 \%$ is the lowest allowing for a detailed global exploration of the phase space within  a -- for us -- reasonably long (O($3\cdot10^4$ $y$) simulation, as it allows for observing a good number (O(40)) of transitions between the $SB$ and the $W$ states. %The estimate of the quasi-potential does not change significantly when considering the output of simulations performed with higher ($\sigma = 20 \%$) intensity of the noise, thus suggesting that the weak-noise approximation is valid. 
We find that the basin of the W attractor is deeper (lower values of the quasi-potential) compared to the basin of the SB attractor. By using Eq.\eqref{eq:tt_distr} and performing an exponential fit of the statistics of average residence times in the two attractors for different values of the noise intensity -- see Fig.~\ref{fig:ocdiff_1_phsp}(c), we obtain the following information on the two local potentials: $\Delta \Phi_{\rm W\rightarrow SB} \approx 700(40)$ and $\Delta \Phi_{\rm SB\rightarrow W} \approx 240(50)$. The good quality of the fit confirms that the weak-noise approximation is valid. 

Another relevant piece of information can be obtained by looking at the paths of the $\rm SB\rightarrow W$ and $\rm W\rightarrow SB$ transitions. In the weak-noise limit, the stochastic average of the trajectories  that manage to escape from either attractor gives the instantonic path  for the portion of trajectory connecting the attractor to an M state, and the relaxation path for the remaining part of the trajectory, which connects the M state to the other attractor. The red (blue) line in Fig.~\ref{fig:ocdiff_1_phsp}(b) indicate the stochastic averages of the $\rm SB\rightarrow W$ ($\rm W\rightarrow SB$) transition trajectories. The procedure for computing the average paths is described in detail in the ESM.

As discussed above, escape trajectories and relaxation trajectories are expected to follow different paths in general nonequilibrium systems. %In previous analyses, while we did confirm that instantons and relaxation trajectories connected attractors and M states, we  were not able to spot a difference in the geometry of the trajectories in the same projection of the phase space considered here, possibly as a result of considering a simpler climate model than PLASIM \cite{Lucarini2019,Lucarini2020}. 
We are indeed able to find such an essential feature of nonequilibrium systems, as clearly detailed in Fig.~\ref{fig:ocdiff_1_phsp}(b). 
In simpler setups with a unique saddle, the crossing point between the red and the blue line must correspond to the position of the M state, see discussion in \cite{Lucarini2019,Lucarini2020}. %, were we to use a different stochastic forcing, we would end up with different instantonic paths for both transitions, but the intersection point can be expected to be the same as here, because it is bound to be on the relevant M state.

Instead,  here the crossing between the two transition paths as observed in Fig.~\ref{fig:ocdiff_1_phsp}(b) is an artifact of looking at that specific two-dimensional projection, as can be noticed  when looking at a three-dimensional projection of the phase space, see Fig.~\ref{fig:ocdiff_1_phsp}(d). The SB$\rightarrow$W and the W$\rightarrow$SB transitions go through two different channels corresponding to two different M states. This marks a major difference with respect to the analysis performed in \cite{Lucarini2019,Lucarini2020}. We have clear indication that in the model used here large-scale currents are present in the phase space, {\color{black} which characterise non-equilibrium conditions; see \cite{Weiss2020} for an application of this concept in a climatic context.}

It is reasonable to ascribe such a  difference to the fact that here we are able to include a large class of processes associated with the transport of water and with its phase changes between solid, liquid, and gaseous forms. Indeed, the hydrological cycle is greatly responsible for the irreversibility of atmosphere \cite{Pauluis2002,Lucarini2009,Lucarini2014} and, at more quantitative level, overwhelmingly contribute to the total entropy production of the geophysical fluids compared to the dissipation of kinetic energy and the turbulent exchange of sensible heat \cite{Goody2000,Lucarini2010b,Boschi2013,Lucarini2014a}. We argue that the lack of a comprehensive treatment of water in the model used in \cite{Lucarini2019,Lucarini2020} leads to an underestimation of the actual entropy production of the system, which makes it closer to equilibrium than the model considered here. According to a statistical mechanics angle, one sees this as associated with the absence (or significant reduction) of probability currents, which are largely suppressed by the presence of a single saddle separating the competing basins of attraction.

Phenomenologically, the presence of clear distinction between the SB$\rightarrow$W and the W$\rightarrow$SB  transition paths explicitly indicates that the global thawing and the global freezing of the planet are fundamentally different processes. %This can be better appreciated by watching the movies portraying the transitions between the two competing states included in the SM.
The thawing proceeds as follows. First, because of persistent positive anomalies of the solar irradiance, the global temperature of the planet grows without much changes in $\Delta T_{EP}$, as the atmospheric circulation is extremely weak and the oceanic transport absent. Then, the equatorial belt starts to melt and, due to the large decrease of the albedo in the equatorial band and subsequent intense warming, $\Delta T_{EP}$ increases substantially -- see the almost vertical portion of the red line in Fig.~\ref{fig:ocdiff_1_phsp}(b). This leads to a strong enhancement of the meridional heat transport performed by the atmosphere and by the ocean, which causes the thawing of the sea-ice at higher latitudes until the sea-ice line reaches very high latitudes compatible with the W climate. 

The global freezing of the planet, instead, proceeds in the following way. The cause of the freezing is, obviously, the presence of a (rare) persistent negative anomaly of the solar irradiance. The reduction of incoming solar radiation has an amplified effect at high latitudes, because of the ice-albedo feedback, leading to an increase of $\Delta T_{EP}$. The increase in $\Delta T_{EP}$ causes a strengthening in the meridional heat transport, which acts as a stabilizing feedback -- see the diagonal portion of the blue line in Fig.~\ref{fig:ocdiff_1_phsp}(b). Nonetheless, if the anomaly in the solar irradiance is sufficiently strong and persistent, the sea-ice line moves equatorward, until the equatorial belt freezes and undergoes further extreme cooling because the albedo becomes very high, leading eventually to a very low value of $\Delta T_{EP}$ in the final SB state.

\subsubsection{Relaxation Modes}
\label{sec:Relaxation_modes_setup_A}
As detailed in the ESM, by constructing a finite-state Markov chain model of the projected $([\langle T_S\rangle],\Delta T_{EP})$ space, one can extract further useful information about the slow dynamics of the system. We study the statistics of the transitions of the state of the system for the case $\sigma =18\%$ on a time scale of 30 days. The dominant eigenvector of the Markov chain is the projection of the invariant measure given in Fig.~\ref{fig:ocdiff_1_phsp}(a). The subdominant eigenvectors describes how a generic initial measure relaxes to the invariant one. %Their integral vanishes, as discussed in Sec.~III of the Supplementary Material, and we normalise them in such a way that the integral of their absolute value is 1. 
{\color{black}We remark that, despite the very severe projection, the Markov chain model features positive metric entropy, which measures the rate of creation of information, and positive entropy production, which unequivocally indicates nonequilibrium conditions and is associated to the presence of currents \cite{Gaspard2004}.}

\begin{figure}[b]
	\centering
	\includegraphics[width=.49\linewidth]{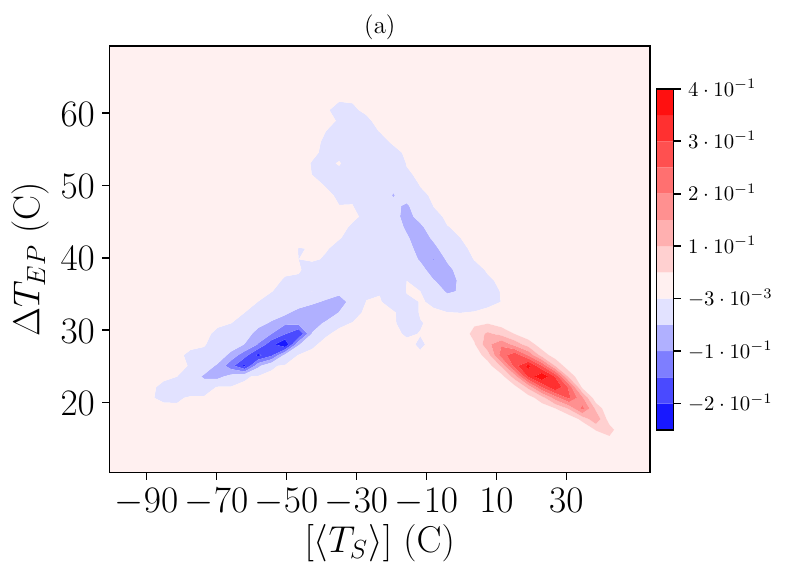}
	\includegraphics[width=.49\linewidth]{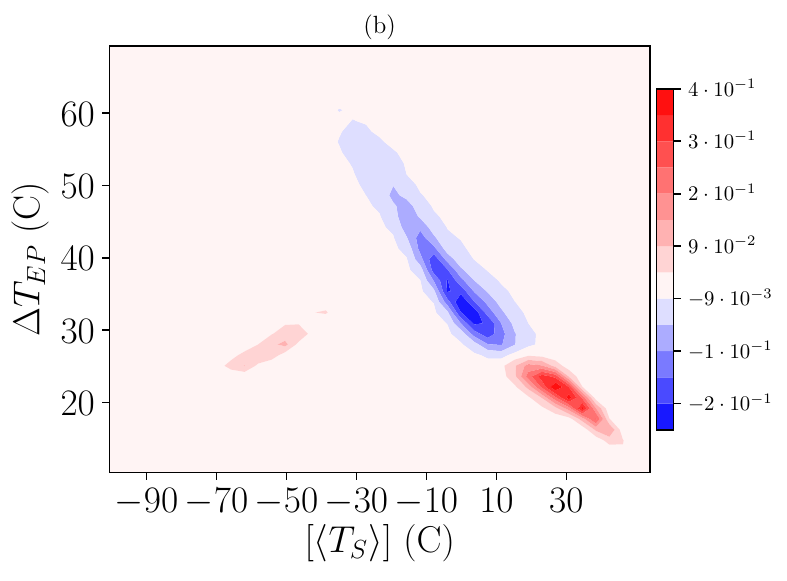} 
	\caption{First two subdominant eigenvectors of the finite state projected Markov operator for setup A and $\sigma=18\%$. (a) First subdominant mode ($\tau\approx 30 y$) describing the transitions between the two competing metastable states; see also a clear signature of persistent cold departures of the system -- within the W basin of attraction --  from typical warm conditions leading to the transitions. %\textcolor{magenta}{This special region within the W basin of attraction is likely to be a left-over of the third attractor realized for setup B}. 
	(b) Second subdominant mode ($\tau\approx 11 y$) describing the low-frequency variability within the W basin of attraction. Note the lack of time-scale separation between these two modes.}
	\label{fig:ocdiff_1_sub} 
\end{figure}

The two leading subdominant eigenvectors of the finite-state Markov chain approximation of the projection of the transfer operator in the $([\langle T_S\rangle],\Delta T_{EP})$ plane for the case $\sigma=18\%$ are presented in Fig.~\ref{fig:ocdiff_1_sub}. Panel (a) describes -- in statistical terms -- the coarse grained, slow process of transition between the two metastable states. One of two peaks is negative and the other one is positive, as the mode describes a zero-sum probability transfer. Additionally, this eigenvector has a very clear signature of persistent excursions of the system in the far cold region of the warm attractor. This might be interpreted as a signature of the preferential regions where transitions between the  $SB$ and $W$ states take place, compare with Fig.~\ref{fig:ocdiff_1_phsp}(b). 

Instead, panel (b) by and large describes the slowest intrawell variability, which takes place in the W basin of attraction: the two closely spaced peaks of opposite sign are on the opposite sides of the peak of the W basin of attraction, with the zero isoline cutting across the peak of the warm attractor; compare with Fig.~\ref{fig:ocdiff_1_phsp}(a). This slow time scale is associated with the process of ice formation and melting. A smaller peak is present in correspondence to the SB basin of attraction, indicating that this eigenvector captures some  $\rm W\rightarrow SB$ escape process; compare with  Fig.~\ref{fig:ocdiff_1_phsp}(b). %Note the different sign of the eigenvector in the vicinity of the $SB$ state with respect to what shown in panel a).   

%As compared with setup A, one discovers that the eigenvector describing the transition between the two metastable states (panel a) has a much shorter life time (about $30$ $y$ compared to $290$ $y$), because the permanence time of the system in the cold attractor is much smaller and, in fact, matches such a life time. Additionally, this eigenvector has a very clear signature of persistent excursions of the system in the far cold region of the warm attractor, which were absent in setup A and probably correspond to shallow regions of the quasipotentials that are a left-over of the third state of the system encountered in setup A. Such excursions are the conducive to transitions to the $SB$ state. The second subdominant eigenvector (panel b), instead, is qualitatively similar to what shown in Fig.~\ref{fig:ocdiff_0_sub}, and we interpret it as describing the slow, large fluctuations of the system inside the $W$ state; note the different sign of the eigenvector in the vicinity of the $SB$ state with respect to what shown in panel a).  

\subsection{Setup B -- Atmospheric-only Large Scale Energy Transport} \label{sec:A third stable climate state}

\subsubsection{The Three Competing Climate States}\label{sec:Climatology of setup A}

Excluding the large-scale heat oceanic transport amounts to removing a very powerful negative feedback, i.e.~a mechanism of stabilization for the climate that efficiently redistributes energy throughout the system. This changes qualitatively the global stability properties of the system compared with the case of setup A. Indeed, in setup B we find three competing climate states, whose basic features are reported in Table~\ref{tab:characteristics}, and we refer to Fig.~2 in the Supplementary Material for further evidence.  One of the climates is the fully-glaciated SB state, which features very low $\Delta T_{EP}$ and extremely low global temperature, close to $-50~^\circ$C. The second climate resembles the W state found in setup A, featuring an above 0~$^\circ$C global temperature, with  $\Delta T_{EP}\approx40~^\circ$C  and roughly 27\% sea ice coverage.  Between the two, lies the -- unexpected and unprecedented for PLASIM -- C state, which is not fully ice covered, and even though it has $[\langle T_S\rangle] \approx -30~^\circ$C, the fact that $\Delta T_{EP}\approx 50~^\circ$C suggests the presence of a warm latitudinal band at subtropical latitudes. The presence of an ice-free latitudinal band has huge implications in terms of habitability \cite{Pierrehumbert2011,Linsenmeier2015}.

\begin{figure}[b]
	\centering		
	\includegraphics[width=0.24\linewidth]{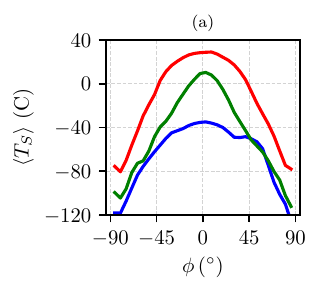}
	\includegraphics[width=0.24\linewidth]{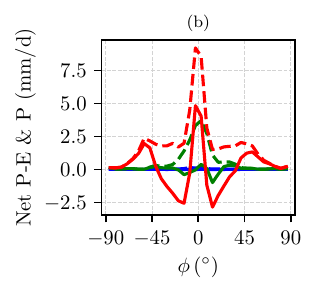}
	\includegraphics[width=0.24\linewidth]{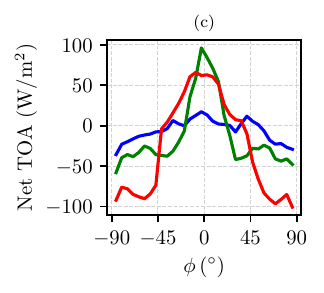}
	\includegraphics[width=0.22\linewidth]{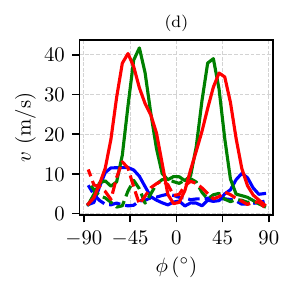}
	\caption{Climatological longitudinally averaged annual mean of (a) surface temperature, (b)  Precipitation minus Evaporation (P-E, solid lines) and Precipitation (P, dashed lines), (c) top of the atmosphere net radiation,  (d) magnitude of zonal wind speed at 300 hPa (solid line) and at 1000 hPa (near surface, dashed lines) versus the latitude $\phi$. Blue lines: SB state. Red lines: W state. Green lines: C state.}
	\label{fig:climatology_setupB} 
\end{figure}
\begin{figure}[t]
	\centering
	\includegraphics[width=0.9\linewidth]{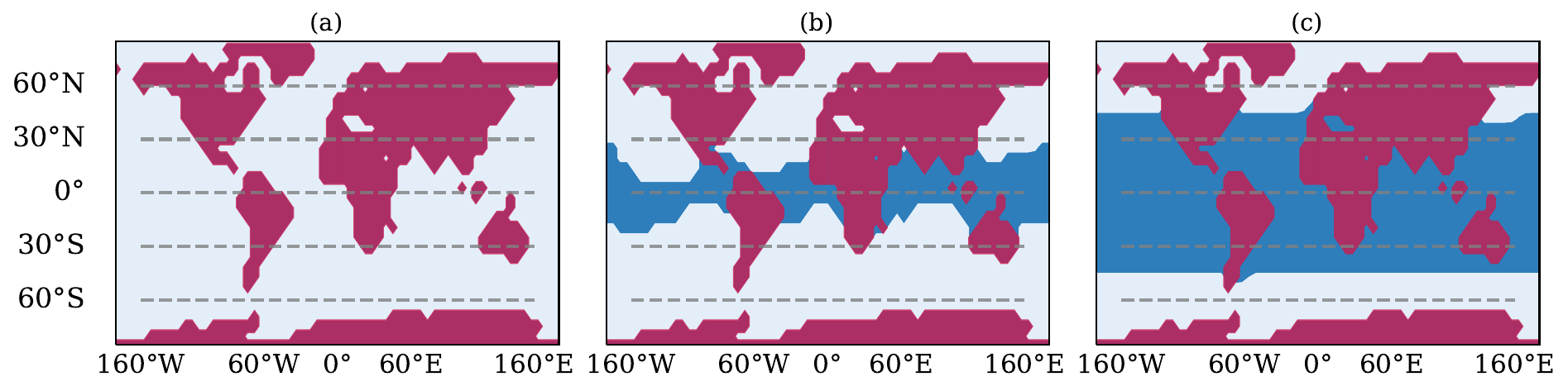}
	\caption{Sea-ice coverage comparison between (a) snowball, (b) cold and (c) warm climates. Note that the W state of setup B has more sea-ice than the W state of setup A.}%\GM{I will change the earth geography to match the grid we used}}
	\label{fig:maps_setupB} 
\end{figure}

In Fig.~\ref{fig:climatology_setupB} we compare the climatology of the three climates (W in red, C in green and SB in blue) \textcolor{black}{resulting from a 40-year average in steady state conditions, in absence of stochastic forcing ($\sigma=0$)}. The SB state is very similar to the one obtained with setup A, as the ocean plays a negligible role in a fully glaciated planet, and will not be further discussed here. The W state is similar with its counterpart in setup A, albeit considerably colder, and, correspondingly, with a weaker hydrological cycle. We can interpret this as resulting from the  ice-albedo feedback. Indeed, the presence of a weaker heat transport towards high latitudes due to removing the oceanic channel of meridional  transport leads to a larger sea-ice surface -- compare Fig.~\ref{fig:maps_setupA}(b) with Fig.~\ref{fig:maps_setupB}(c) -- which contributes to lowering the planetary albedo, thus enhancing the input in the energy channel at TOA. Due to the Boltzmann radiation feedback, the steady state must then be characterized by a lower average temperature compared to setup A. Finally, the presence of larger temperature differences between high and low latitudes lead to a stronger atmospheric variability, as baroclinic conversion is more efficient and can draw from a larger reservoir of available potential energy. This is associated with a stronger Lorenz energy cycle compared to setup A, see Table~\ref{tab:characteristics}; see a discussion of the climatic effects of modulating the meridional oceanic heat transport in the W state in \cite{Knietzsch2015}.

Fig.~\ref{fig:climatology_setupB}(a) shows the climatology of the zonal mean surface temperature. We remark that in the C state the subtropical band  $[-20^\circ N,20^\circ N]$ features above-freezing temperature, while lower temperatures, and correspondingly, prevailing sea-ice is present at higher latitudes, as shown in Fig.~\ref{fig:maps_setupA}. {\color{black} Despite PLASIM's simplified dynamics,} the C state shares features of the previously mentioned  Slushball state \cite{Hyde2000} and, especially, of the Jormungand state \cite{Abbot2011}, where the presence of ice-free equatorial band is associated with the dynamics of continental ice sheets and of the interplay of sea-ice cover, surface albedo, and atmospheric circulation, respectively. Figure~\ref{fig:climatology_setupB}(b) shows the annual mean budget of the precipitation minus evaporation rate (P-E; solid lines) as well as the annual zonally averaged precipitation (dashed lines). 
%The SB climate is almost entirely dry, as a result of the fact that the very low temperature of the atmosphere permits the presence of nothing but an extremely small amount of water vapour, because of the constraint posed by the Clausius-Clapeyron relation. The W climate has the familiar maximum of precipitation in the equatorial belt and secondary peaks in the mid-latitudes, resulting from convective precipitation and synoptic disturbances, respectively. The net PE describes the familiar scenario of new water vapour transport from the tropics into the equatorial belt and into the mid-latitudes \cite{Peixoto1992}. 
The C state features an intense precipitation in the equatorial belt, driven by the strong convection occurring there, but the P-E field indicates that the water vapour is recycled and no large scale transport takes place, as opposed to the W state.

Figure~\ref{fig:climatology_setupB}(c) shows the zonally averaged net TOA energy budget. One can infer that the meridional atmospheric enthalpy transport has comparable intensity in the W and C climates, yet the peaks of the transport -- indicated by vanishing values of the TOA budget \cite{Peixoto1992,Trenberth2009,Lucarini2011}- are confined to lower latitudes in the latter case. This indicates a vigorous heating realized at $\approx \pm 30^oN$. Correspondingly, the jet stream for the C state is located at lower latitudes compared to the W climate (panel d), while it is more intense, as the local meridional temperature gradient throughout the atmosphere is larger. This corresponds to a large temperature difference between low and high latitudes at surface, see Table~\ref{tab:characteristics}. %for the . %Instead, the meridional enthalpy transport and the zonal circulation of the SB state are, as well know, extremely weak, corresponding to the presence of very weak meridional temperature gradients.  \cite{Lucarini2010,Lucarini2017a}. The SB state features as very weak Lorenz energy cycle ($\approx 1.2 W m^{-2}$), as the presence of a weak meridional temperature gradient leads to the presence of a scarce reservoirs of available potential energy and shuts down almost entirely the mechanism of baroclinic instability. 
%The vigorous circulation of the W state corresponds to a very powerful Lorenz energy cycle ($\approx 4.7 W m^{-2}$). 

Finally, the C state features a strong Lorenz energy cycle ($\approx 4.0 \rm W m^{-2}$), thanks to the presence of such large meridional temperature gradients which correspond to a large reservoir of available potential energy that can be converted to kinetic energy by baroclinic instability. The intensity of the Lorenz energy cycle of the C state is especially remarkable given that the atmospheric circulation is relatively weak poleward of $50^\circ$ latitude.

\subsubsection{Noise-induced Transitions}\label{sec:Noise induced transitions}

\begin{comment}
As soon as the system has reached a nonequilibrium steady state it is contained in one of the three discussed attractors, and it is impossible to escape deterministically. As stated earlier, though, transitions among the attractors can be induced by noise. In Fig.~\ref{fig:ocdiff_0_phsp}(a) we present the projection of the invariant measure on the reduced phase space spanned by $[\langle T_S\rangle]$ and $\Delta T_{EP}$, while in Fig.~\ref{fig:ocdiff_0_phsp}(b) we approximately estimate the corresponding quasipotential using Eq.~\eqref{eq:stationary_distr} as 
\begin{equation}\label{eq:quasipotential}
\Phi([\langle T_S\rangle],\Delta T_{EP}) \sim - \frac{\sigma^2}{2}\log \rho_\sigma([\langle T_S\rangle],\Delta T_{EP}).
\end{equation}
Notice that we subtract the absolute minimum from $\Phi([\langle T_S\rangle],\Delta T_{EP})$ to shift to zero the energy level of the deepest potential well. In both cases $\sigma = 12 \%$, which is the weakest noise used for setup A.
\end{comment}

\begin{figure}[ht!]
	\centering
\includegraphics[width=.49\linewidth]{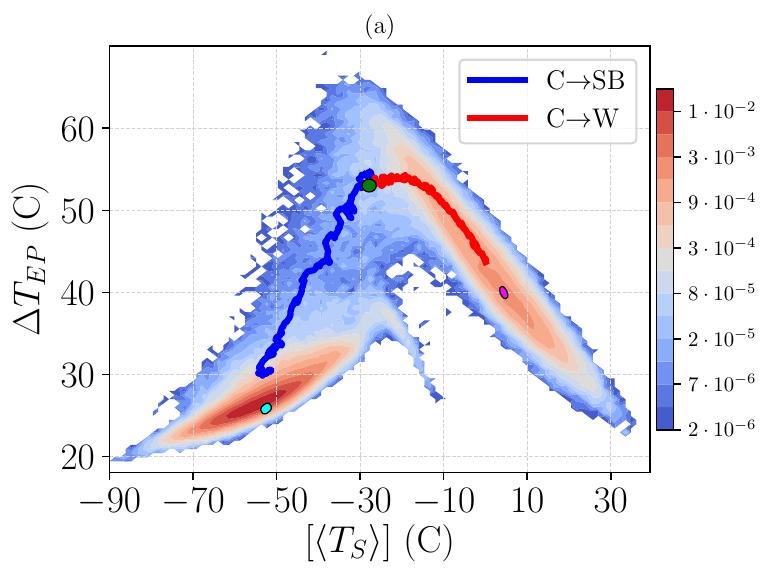}
	\includegraphics[width=.49\linewidth]{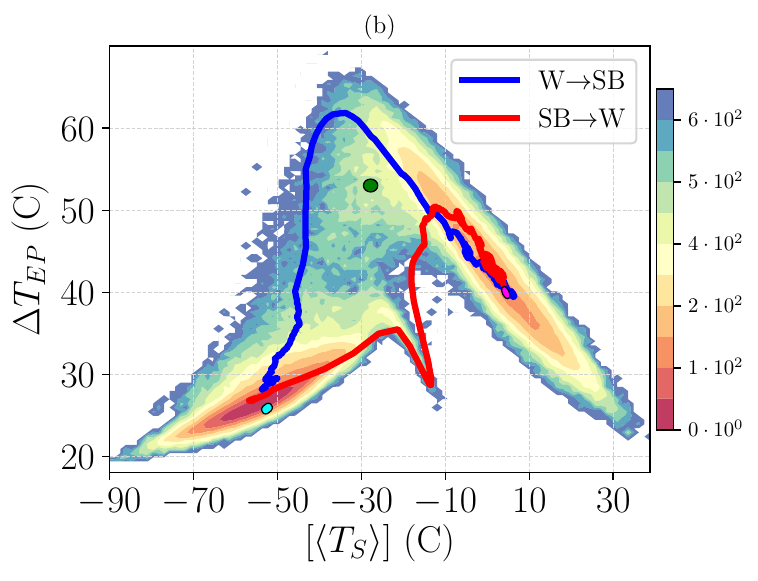}
	\caption{ (a) Projection of the invariant measure of the system in the reduced phase space composed of the 30-day averaged global temperature versus 30-day averaged Equator to pole temperature difference for setup B and $\sigma=12 \%$. The stochastically averaged escapes from the C state with $\sigma=6\%$ are also included. (b) The corresponding quasipotential, shifted so that the global minimum is set to 0. The blue and red lines correspond to stochastically averaged transition paths  for the W$\rightarrow$SB and SB$\rightarrow$W transitions. The trajectories are completely different, depending on the direction of the transition. The  ellipses indicate the location of the deterministic attractors corresponding to the SB state (cyan), the C state (green) and the W state (magenta). } 
	\label{fig:ocdiff_0_phsp} 
\end{figure}

The presence of three instead of two deterministic attractors makes setup B considerably more complex than setup A; for example now the existence of extra M states connecting SB with C and W with C has to be taken into account, on top of those connecting SB with W already seen in setup A. Figure~\ref{fig:ocdiff_0_phsp}(a) shows the projection of the invariant measure in the reduced phase space given by $([\langle T_S\rangle],\Delta T_{EP})$ obtained for $\sigma=12 \%$, while in Fig.~\ref{fig:ocdiff_0_phsp}(b) we show the corresponding estimate of the quasipotential. We remark that in setup B a lower noise intensity is needed to excite transitions with frequency comparable to what obtained in setup A, for the basic reason that we are missing the global stabilizing feedback given by the ocean heat transport. This corresponds to having weaker diffusion in the Fokker-Planck operator describing the evolution of probabilities. The location of the deterministic attractors is shown with ellipses of different color, where magenta, green and cyan correspond to W, C and SB climate states, respectively.

%The ellipses are estimated by initiating a few relaxation trajectories, by now keeping fixed the solar constant to $S^* = 1365 \, \textrm{W}/\textrm{m}^2$, in a selected region of the phase space $([\langle T_S\rangle],\Delta T_{EP}) = (25\pm5,40\pm5)$.  After evolving the relaxation trajectories for a few thousand years, to ensure stability, we measure the mean and standard deviation of the observable's timeseries. Finally, the center of the ellipse is placed according to the mean of the observables, and its height and width according to their standard deviation, while for visualization purposes we magnify the ellipse by a factor of 2 or 4.

The location of the C state is not directly visible in the projected invariant measure or in the quasi-potential, in the form of a local maximum and minimum, respectively. The operation of performing a projection to such a low-dimensional space is mainly responsible for such a loss of information.  {\color{black} This issue is addressed specifically in Sec.~\ref{datascienceresults}.} Additionally, as we shall see below, the third attractor corresponds to a much shallower local minimum of the quasi-potential compared to the W or SB states. As a result, the C local minimum is washed out when considering  a noise intensity of $\sigma = 12 \%$, and it is hard to keep track of orbits persisting significantly near C, see Eqs.~\eqref{eq:stationary_distr}-\eqref{eq:tt_distr}. This implies the presence of an additional scale relevant for understanding the multistability of the system, along the lines of what depicted in Fig.~\ref{fig:multiscale}.

\begin{figure}[h!]
	\centering
	\includegraphics[width=.49\linewidth]{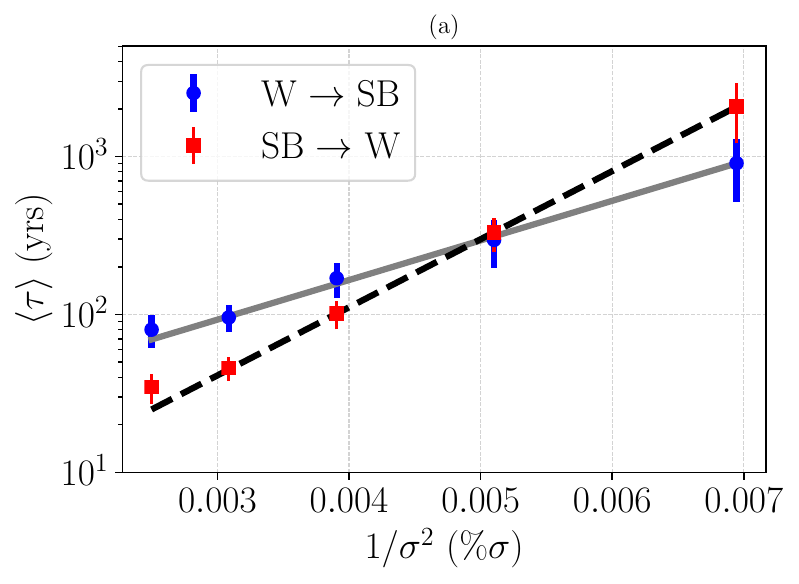}
	\includegraphics[width=.49\linewidth]{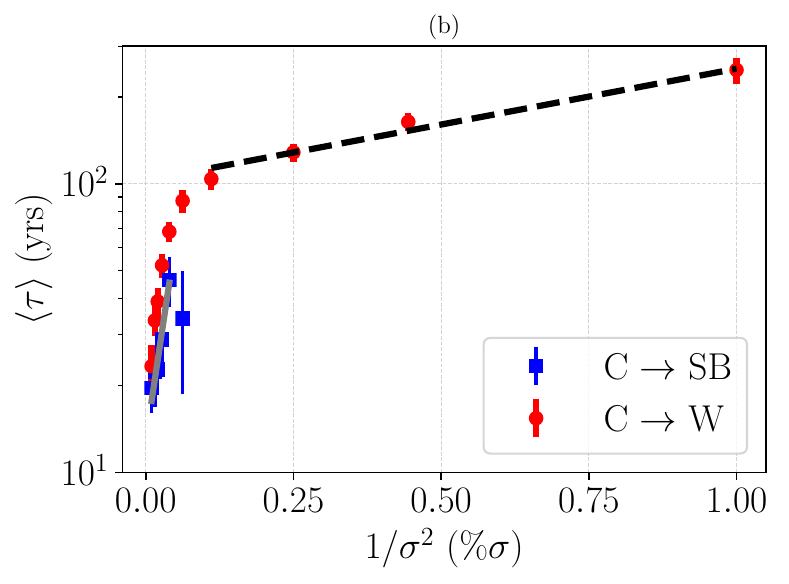} 
	\caption{Average escape time versus the inverse squared $\%\sigma$ in setup B. (a) Comparison of W$\rightarrow$SB (filled blue circles) and SB$\rightarrow$W (filled red squares) and corresponding exponential fit, gray straight and black dashed lines. (b) Comparison of C$\rightarrow$SB (filled blue squares) and C$\rightarrow$W (filled red circles) and corresponding exponential fit, gray straight and black dashed lines. The fitting expression is given by Eq.~\eqref{eq:tt_distr}.}
	\label{fig:escapes_atr} 
\end{figure}

As mentioned above, the presence of ocean diffusion triggers the ice-albedo feedback in a direction that favours warming. Accordingly, in setup B, the minimum of the quasipotential corresponding to the SB state is deeper than the one corresponding to the W state. This can be seen in Fig.~\ref{fig:escapes_atr}(a), where the $SB\leftrightarrow W$ mean escapes times $\bar{\tau}_\sigma$ are presented as a function of the inverse squared noise amplitude. Using Eq.\eqref{eq:tt_distr}, we obtain the following estimates for the depth of the local quasi-potentials: $\Delta \Phi_{\rm W\rightarrow SB} \approx 290(10)$ and $\Delta \Phi_{\rm SB\rightarrow W} \approx 500(10)$.  As opposed to setup A, in setup B the pre-exponential factors of the expectation value of escape times is vastly different. Note that,  neglecting the C state, the population of the SB and W state is inversely proportional to the corresponding escape times. As a result, despite being associated to a shallower local minimum of the quasi-potential, the fraction of population in the W state is larger when considering relatively strong noise intensity, whereas eventually, the SB state dominates in the  weak-noise limit. Despite the profound dynamical differences between setup A and B, the estimates of the instantonic and relaxation paths between the SB state and the W state are qualitatively similar; compare Figs.\ref{fig:ocdiff_1_phsp}(b) and~\ref{fig:ocdiff_0_phsp}(b). Furthermore, the interpretation of the different physical mechanisms controlling the SB$\rightarrow$W and W$\rightarrow$SB transitions paths for setup B is fundamentally the same as for setup A.  

The more complex geometry of the phase space of setup B is  made apparent by the fact that (see the movies included in the ESM), the transitions between the W and SB states can be either direct or, instead, the paths deviate considerably as the orbit is temporarily trapped near the C state. Such a trapping is always extremely short-lived compared to the other relevant time scales associated with the transition between the two other metastable states.

The next step is to provide a characterization of the quasi-potential near the C state, and, specifically, to  estimate the C$\rightarrow$SB and C$\rightarrow$W barriers for the local quasi-potential. %In order to get around the curse of the multiscale nature of the multistability, which entails that no selection of noise is suitable for getting a complete picture of the system, because it is either too large or too weak for discovering some features of the quasi-potential, we proceed as follows.
Following \cite{Lucarini2019}, we investigate the escape process from the C state by considering a large number of trajectories initialized in the deterministic C attractor and apply a weaker random forcing with $\sigma= (1\% - 10\%)$. We then collect the statistics of escape times and keep a separate track for trajectories ending up in the W versus in the SB state through the corresponding M states. Using Eq.~\eqref{eq:tt_distr}, we are able to estimate the two quasi-potential barriers $\Delta \Phi_{\rm C\rightarrow SB}$ and $\Delta \Phi_{\rm C\rightarrow W}$. We see in Fig.~\ref{fig:escapes_atr}(b) that  $\Delta \Phi_{\rm C\rightarrow SB} \approx 16(2)$ (blue filled squares) is about one order of magnitude smaller than the W$\leftrightarrow$SB barriers. Interestingly, the energy barrier $\Delta \Phi_{\rm C\rightarrow W} \approx 0.45(4)$ (red filled circles) turns out to be much smaller than $\Delta \Phi_{\rm C\rightarrow SB}$, which explains why bellow a certain noise level, i.e.~$\sigma \approx 4\%$ we practically get no transitions towards the SB attractor, with all escape trajectories ending in the W basin of attraction. Also, for the  C$\rightarrow$W transitions, we clearly observe from Fig.~\ref{fig:escapes_atr}(b) that for $\sigma$ larger than $\sigma \approx 5\%$ there is a different scaling that can be attributed to the prefactor in Eq.~\eqref{eq:tt_distr}, which indicates that the weak-noise limit is not achieved for these values of $\sigma$ for these escape processes. Further comments on the escape from the C state can be found in the ESM.

\begin{figure}[t!]
	\centering
	\includegraphics[width=.49\linewidth]{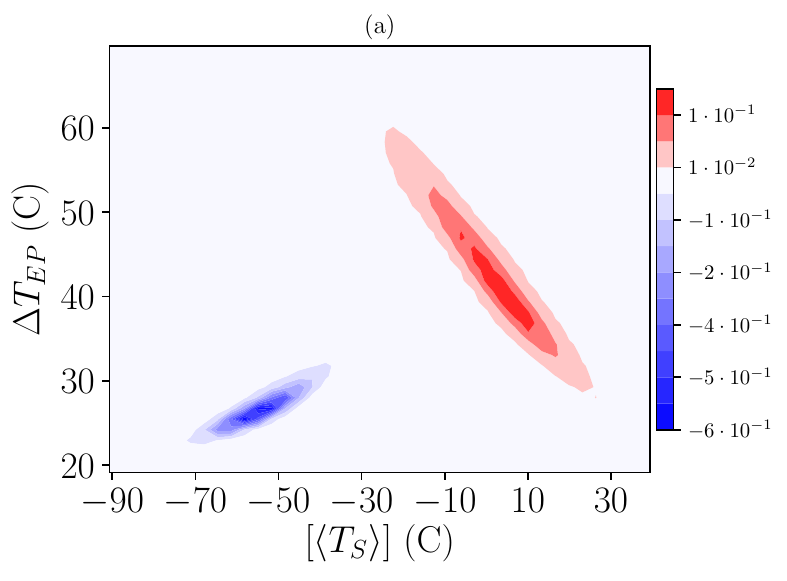}
	\includegraphics[width=.49\linewidth]{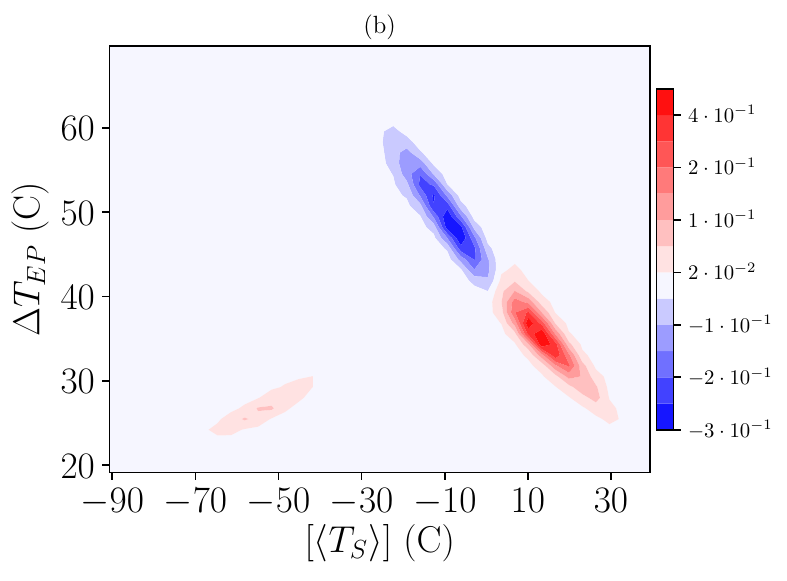} 
	\caption{First two subdominant eigenvectors of the finite state projected Markov operator for setup B and $\sigma=12 \%$. (a) First subdominant mode ($\tau\approx 290 y$) describing the transitions between the two main competing metastable states. (b) Second subdominant mode ($\tau\approx 10 y$) describing the low-frequency variability within the warm attractor. Note the lack of any signature of the cold attractor, whose lifetime is much shorter than 10 years at noise level $\sigma=12 \%$.}
	\label{fig:ocdiff_0_sub} 
\end{figure}

\subsubsection{Relaxation Modes}
Finally, we study the two subdominant eigenvectors of the finite-state Markov chain approximation of the projection of the transfer operator in the $([\langle T_S\rangle],\Delta T_{EP})$ plane for the case $\sigma =12 \%$, see Fig.~\ref{fig:ocdiff_0_sub}. As in setup A, the Markov chain model features positive metric entropy and positive entropy production. We get a broad agreement with the results of setup A also in terms of interpretation of the meaning of the eigenvectors, but a more clear separation of scales between the two corresponding eigenvalues is evident in this case. In Fig.~\ref{fig:ocdiff_0_sub}(a) the first subdominant eigenvector has a much longer life-time of approximately 290 years, which matches the life time of the SB state. Because of such a long time scale, and of the fact that the transition time is very short compared to the residence time, %, actual time it takes the system to cross over the M state is small compared to the permanence in either the SB or M state, 
we lose any feature of the transition path, as opposed to setup A. The eigenvector shown in Fig.~\ref{fig:ocdiff_0_sub}(b) has a life-time of about 10 years and portrays the low-frequency variability in the W basin of attraction, which can lead to occasional transitions towards the SB state; compare the $\rm W\rightarrow SB$ transition path in Fig.~\ref{fig:ocdiff_0_phsp}(b). We find no signature of the presence of the C state, whose life time is much smaller than 10 years for this level of noise. These eigenvectors further clarify that for this level of noise the C state is almost entirely washed out.

\subsection{Automatic determination of the metastable states}\label{datascienceresults}
%The previously discussed results have been obtained using the -- methodologically standard -- top-down approach based on the careful use of targeted numerical simulations, differing for the chosen intensity of stochastic forcings or altogether neglecting it.
The basic issue we want to address now is that, while in Fig.~\ref{fig:ocdiff_0_phsp} the SB and W state clearly appear as corresponding to local maxima of the projected invariant measure, this is not the case for the C state, in this as well as in many other 2D projections we have tested. Indeed, it has been  impossible with the tools developed so far to find any direct evidence of the C state in the stochastic integrations. As described in Sect.~\ref{sec:Climatology of setup A}, the discovery of the C state has been serendipitous and based on the exploration of the phase space via forward deterministic simulations. We next show what can be obtained by applying the suite of data driven methods \cite{faccoID,rodriguez2018computing,science_DPA} presented in Section~\ref{sec:topography} to 
%about the structural properties of the climate by treating 
the output of some given numerical simulations taken as pseudo-observations of an in principle unknown model. %, and applying a suite of data driven methods \cite{faccoID,rodriguez2018computing,science_DPA} presented in  Section~\ref{sec:topography}.

We first consider a numerical integration of the model in setup B lasting $6\cdot10^4$ years and performed with $\sigma=12 \%$. From the complete trajectory of $O(10^5)$ d.o.f.~recorded with having temporal resolution of one time step, we construct a severely coarse-grained version of the phase space by a set of 30-day averaged air temperatures measured every 10 months (hence, decimated with respect to the standard 30-day averaged dataset in previous sections) at three different pressures (300, 500 and 1000 hPa) and 32 different latitudes, for a total of $n=96$ variables.  The quasipotential as a function of these variables  is, in principle, a 96-dimensional function, which  cannot be visualized or estimated in a simple manner. 

\begin{figure}[t!]
	\centering
	\includegraphics[width=0.49\linewidth]{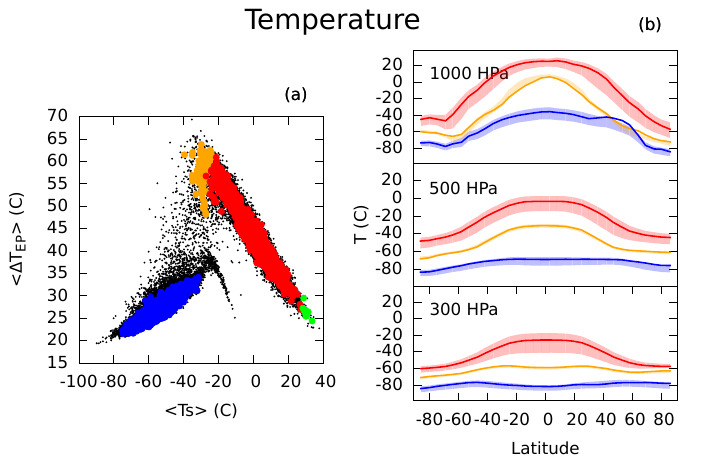}
	\includegraphics[width=0.49\linewidth]{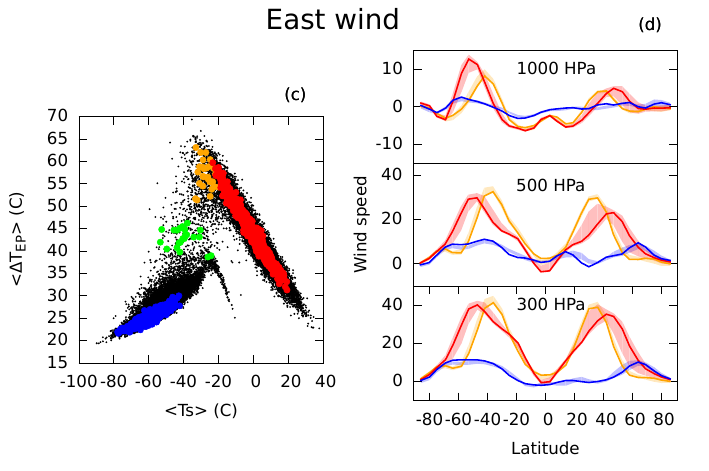}
	\caption{The topography of the quasipotential  in high-dimensional spaces. Panel (a) and (b): the analysis is performed for setup B and  $\sigma=12 \%$  in the coordinate space of the air temperature at three different pressures (300, 500 and 1000 hPa) at 32  latitudes between -86$^o$ and 86 $^o$ (96 variables). Panel (c) and (d): the analysis is performed on the time series of the zonal and the meridional wind in the same latitudes and pressures (192 variables). Panel (a) and (c) portray the estimated basins of attraction of the quasipotential, which are represented in the $([\langle T_S\rangle], \Delta T_{EP})$ projection. The core sets of each attractor are colored in blue (SB state), orange (C state) and red (W state). The green points are the core set of spurious attractors found by the algorithms in both setups. The black points are configurations which do not belong to any core state. Panel (b) and (d) portray the average value plus/minus one standard deviation of the variables, restricted to the core sets of the SB (blue), W (red) and C (orange) states, as light shaded area. In comparison, the time averages of the same variables, computed for the corresponding deterministic attractors, is shown in dark solid lines of the same color, respectively.
		The meridional wind is not shown.
	}
	
	\label{fig:clusterT} 
\end{figure}

By using the approach outlined in Sec. \ref{sec:topography}, we study the topography of this function. We first estimate the intrinsic dimension of the manifold containing the data, which turns out to be $\sim 11$,  significantly smaller than the number of variables {\color{black}\footnote{Note that we should not in any way interpret this number as representative of the actual effective dimension of the attractor of the climate system, because the coarse graining procedure applied in space and time filters out almost entirely the dynamics -- which is prevalent in this climate model as well as in reality -- occurring over time scales shorter than one season and featuring longitudinally symmetric structure \cite{Ghil2020}.}}.
This number is approximately scale invariant: indeed the estimated value does not change significantly if the data set is significantly undersampled. % (see {\color{black}Supp Fig. xx}). 
Since the intrinsic dimension of the embedding manifold is relatively low and well-defined, one can estimate  the quasipotential  $\Phi_t$ in each time frame $t$ using Eq.~\eqref{eq:F_PAk}, without defining explicitly the $\sim$ 11 coordinates mapping the manifold. Using these estimates, one finds the attractors, which correspond to the local minima of $\Phi$. With a statistical confidence level of 99\%, corresponding to $Z=2.576$, we find 4 states, with a core population of 39171, 12099, 112 and 11 frames respectively. The configurations corresponding to the four minima of $\Phi$ were then evolved without stochastic forcing in order to obtain the corresponding asymptotic states, While the first three states are in the basin of attraction of the SB, W, and C attractors, correspondingly, the fourth state is found to be unstable, as it forward evolution converges to the W attractor. This indicates that the fourth state is an artifact of finite sampling, or of the variations of the $Z(x)$ (see Eq.~\eqref{eq:stationary_distr}) which in the estimate of $\Phi_t$ are neglected. The configurations assigned to the core set of the three remaining states are represented in Fig.~\ref{fig:clusterT}(a) in the same projection used in Fig.~\ref{fig:ocdiff_0_phsp}. In this projection the C and W states strongly overlap, and no barrier is visible between the two. 

In Fig.~\ref{fig:clusterT}(b) we plot the average and the standard deviation, estimated for the core set of each state, of the 96 air temperature variables used in the analysis. Note that such average values agree remarkably well with the time-averages one obtains by considering the corresponding deterministic attractors, represented as continuous lines in Figure~\ref{fig:clusterT}. Remarkably, the  distributions are significantly well separated for almost all the variables. This demonstrates that the W and C state %, which are overlapping in the two-dimensional projection of panel a, 
are indeed non-overlapping in the 96-dimensional space of these variables. This also shows that the data-driven approach presented here %that feeds upon an individual trajectory of the stochastically perturbed system 
is able to reconstruct accurately the statistical properties of the competing deterministic metastable states. %The fact that the error bars computed for the reconstructed states are centered to a very good approximation on the corresponding values obtained for the deterministic attractors also indicates that the core sets in high-dimensional space are approximately Gaussian.}

We then repeated the exercise by considering the $n=192$ variables describing the 30-day averaged meridional and zonal wind at the same latitudes and pressure levels as before. The intrinsic dimension of this dataset is $\sim 16$, slightly larger than for the other variables. In this space, at a statistical confidence of $99\%$ the algorithm is able to detect only two states, the W and the SB states. At a $98\%$ confidence the C states appears (orange points in Fig.~\ref{fig:clusterT}(c)), together with another state, represented in green. The latter state turns out to be spurious, since simulations performed with $\sigma=0$ starting from the estimated minimum rapidly converge to the SB state. In this space the C state is much more similar to the W state, as illustrated in Fig.~\ref{fig:clusterT}(d): the average zonal wind differs significantly only in the mid-latitudes of the Southern Hemisphere at all levels and in the mid-latitudes of the Northern Hemisphere only at 500 hPa. Note also in this case the excellent agreement obtained with the average statistics computed for the corresponding deterministic attractors. 

\begin{figure}[t!]
	\centering
	\includegraphics[width=1\linewidth]{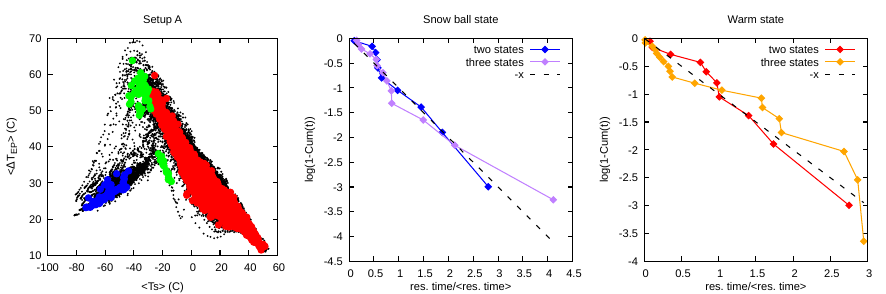}
	\caption{The quasipotential and the residence times for setup A and $\sigma=18 \%$. Panel (a): the states obtained analyzing the coordinate space of the air temperatures at three different pressures   at  32  latitudes (the same variables used in the analysis in Fig.~\ref{fig:clusterT}-a). The analysis is performed with Z=5. At higher statistical significance the green state disappears. Panel (b) and (c): the empirical cumulative distribution  $C(x)$ of the normalized escape time $x=t_{\text{esc}}/<t_{\text{esc}}>$, where $ <t_{\text{esc}}>$ is the  average of all the observations. Panel (b):   the SB state. Panel (c): the W state. Blue and red lines: the green state in panel a is not considered meaningful. Purple and orange lines: the green state is considered meaningful. The dashed black lines correspond to the ideal case in which $x\sim \exp(-x)$, and therefore $\log(1-C(x))=-x$. }
	\label{fig:restime}
\end{figure}

We also performed the same analysis on a simulation evolved for 32780 years using the model in setup A and with $\sigma =18\%$. At high statistical significance, we detect two states corresponding to the W and the SB climates, described in Sect.~\ref{sec:Climatology of setup A}. At lower statistical significance other states appear. An example of an analysis is shown in Fig.~\ref{fig:restime}(a), where the green state approximately seem to occupy approximately the same region as the C state found in setup B, see Fig.~\ref{fig:clusterT}(a). However, the distribution of the air temperature variables in this state differs significantly from the C state in setup B (not shown). % {\color{black}. %(see Supp fig xxx)}.  
Indeed, this state is not an attractor, as it evolves towards the W state if one removes the stochastic forcing. %configuration of lower $\Phi$ belonging to this state, if evolved with deterministic dynamics by setting $\sigma=0$, converges 
rapidly to the W state. 
The dynamics of an ensemble of trajectories initiated near the green dots is by and large controlled by two subdominant eigenvectors depicted in Fig.~\ref{fig:ocdiff_1_sub}(a-b).

These results indicate that our approach allows identifying the correct metastable states of a complex high-dimensional dynamic model, but these states come with an uncertainty, which partially derives from statistical errors, and partially from the approximations intrinsic in the quasipotential estimator, which neglects the preexponential factor $Z(x)$. Finally, an error is introduced by the correlation between the frames, which are generated by a dynamic model and sampled with a time lag of a few months. 
However,  one can rather  straightforwardly recognize the spurious states, even without performing a simulation at $\sigma=0$, by estimating, on  the same trajectories which brings to their identification, the probability distribution of the first escape times. 
This distribution is estimated by  assuming that the system performs a transition between two states when it visits a core configuration belonging to a state which is different from the state of the last core configuration visited in the past \cite{schutte2011markov}. In  this manner, one splits the trajectory in  segments, each labeled with a different state, whose length is an estimate of the escape time $t_{\text{esc}}$. If the set of states defines (at least approximately) a Markov model, $t_{\text{esc}}$ should be exponentially distributed. In Figs.~\ref{fig:restime}(b-c) we plot a function of  the empirical cumulative probability distribution of $t_{\text{esc}}$ which, if $t_{\text{esc}} \sim \mathrm{Exp}$, should coincide with the black dashed lines. If one considers as meaningful also the green state in Fig.~\ref{fig:restime}(a) one obtains a set of   $t_{\text{esc}}$ from the W and the SB state whose distribution significantly deviates from an exponential (purple and orange lines in panels b and c). If instead one does not consider the green state as meaningful, the distribution of the escape times from the W and SB state is almost perfectly exponential (blue and red lines), as far as one can judge from the relatively small number of transition events observed in the trajectory.
This analysis indicates that our approach allows identifying the \emph{correct} metastable states of the system even from relatively short trajectories, in which only  $\mathcal{O}(10)$ transitions are observed. The states can be identified in a fully unsupervised manner, analyzing only the trajectory or by running short relaxation dynamics with $\sigma=0$.

\section{Conclusions}\label{sec:Conclusions}
Achieving a deeper understanding of the nature of the Earth's multistability and related tipping points is one of the key contemporary challenges because it is essential for better framing the co-evolution of climatic conditions and of the biosphere throughout the Earth's history, and, in the present context, for better constraining the current planetary boundaries through a careful examination of the safe operating space for Humanity \cite{Rockstr2009}.

{\color{black} Systems undergoing stochastic dynamics and featuring competing multistable states can be effectively described by taking advantage of the formalism of the quasi-potential landscape, which generalizes the notion of the free energy to nonequilibrium systems. Local minima in the quasipotential describe competing metastable states, and are separated by local maxima and saddles -- M states -- that define possible gateways for transitions.  To demonstrate our framework in the case of the climate we employ} two versions of an open source climate model, PLASIM, which has an appropriate mix of precision, flexibility, and efficiency in simulating the present climate as well as very exotic climatic conditions. The first version (setup A) features a simplified but meaningful representation of the oceanic energy transport from low to high latitudes, whereas in the second one (setup B) large scale energy transport is provided solely by the turbulent atmosphere. Setup A demonstrates the well-known competing climatic states corresponding to the present warm (W) conditions and the so-called snowball (SB) climate. Setup B, instead, contains an unexpected additional intermediate stable climate (C) where the sea is partially ice-free in the equatorial band. The lack of a powerful mechanism of energy redistribution across the climate makes this additional state possible. {\color{black}Despite PLASIM's relative simplicity,} the C state should not be regarded as a pure mathematical curiosity corresponding to a pathological solution: exotic climate states rather similar to the C state obtained here have been obtained in other climate models and are deemed extremely relevant in  paleoclimatic terms because they provide a scenario able to explain the survival of life during the Neoproterozoic glaciations. %\textcolor{red}{cite!}. 

The phase space of the model can be explored when stochastic forcing -- here in the form of a yearly fluctuating solar irradiance -- is introduced, leading to transitions between the competing metastable states. %Indeed, many other sources of stochasticity are present in the real climate system. Moreover, in order to be able to explore the full phase space of the system within a reasonable time, we are compelled to consider a much stronger noise than what is deemed physically realistic for our planet, yet this has actually little impact on our main conclusions, because of the mathematical structure of the problem.
We compute the quasipotential function, which describes, on the one side, the invariant measure of the system and, on the other side, in its local version, controls the probability of transition of the stochastically forced trajectory from one to another basin of attraction. We are able to estimate in both setups the optimal escape paths -- the instantons -- and the corresponding relaxation trajectories linking the W and SB states, and are then able to verify the nonequivalence between the two, which is an essential feature of nonequilibrium properties. %The intersection between instantons escaping the SB and the W attractors meet at the M state, the saddle contained  in the basin boundary that separates the two corresponding basins of attraction. 

Instantons describe how transitions take place in the zero-noise limit and are more of a mathematically elegant construction than a physically relevant object in our investigations, as we need to consider noise of moderate yet non-negligible intensity in order to observe reasonably frequent transitions between the SB and W attractors. Additionally, studying the transfer operator in a suitably projected space sheds light on how the system relaxes to its invariant measure. We are able to find clear evidence of both interwell relaxation processes, which describe transitions between competing metastable states, and are the noisy version of instantons, and intrawell relaxation processes, which would conventionally be labelled as ultralow frequency variability within the W state associated with large scale melting and thawing of sea ice and corresponding large temperature fluctuations. 

A nontrivial result we obtain is that the instantons escaping the SB and the W attractors do not meet at one of the M states separating the two corresponding basins of attraction. This can be best appreciated visually by watching the videos included in the ESM. In fact, the transitions take place through two separate saddles. This has two important implications a) the dynamics on the basin boundary is, by itself, multistable; and b) one has large-scale nonvanishing currents in the phase space. This is a strong signature of the nonequilibrium nature of the system. The existence of separate paths for the SB-to-W and W-to-SB states marks a relevant difference with previous studies. The presence of more evident macroscopic signature of nonequilibrium conditions  can be attributed to the presence in this model of an active hydrological cycle, which is the major agent of entropy production in the climate system. %The detection of the saddles using the classical edge tracking algorithm poses additional challenges compared to the standard case, because the dynamics on the basin boundary is, by itself, multistable.}

%The investigation of the properties of the C state in setup B clarifies the multiscale nature of multistability of the climate system. 
The C state in setup B  corresponds to a comparably shallower minimum of the quasipotential, which can be explored only considering significantly weaker noise than needed to explore globally the phase space of the system. We discover that the most natural, preferential escape route from the C state is towards the W state. The C state is only barely metastable, as even internally generated noise of the numerical discretization can destabilize it, even if only rarely and over ultra long time scales. The position in phase space of the C state and its properties indicate that it is likely that the C state is the leftover of the M state between the SB and W climate obtained as we progressively switch off the horizontal diffusivity of the ocean, because this leads to a less efficient redistribution of energy in the system,

We have complemented the top-down approach based on numerical modelling  with bottom-up data-driven methods that allow for the automatic detection of the competing metastable states from the analysis of a single long stochastic trajectory and to reconstruct the quasi-potential in arbitrarily high-dimension. %\textcolor{red}{This is particularly important for studying the dynamics: the less aggressive dimensional reduction is, the more one can assume that the noise in the process representation is a memory-less process, and use Eq.~\eqref{eq:tt_distr} to estimate the rates.} 
Using this approach we have been able to reconstruct the dynamical landscape of the climate model in both setups and gain a better understanding of how transitions between the competing metastable states occur. Remarkably, by suitable averaging over many realizations, we have been able to reconstruct the climates of the competing (deterministic) metastable states. 

\subsection{Outlook: Multiscale Multistability}
\label{sec:Conc_outlook}

{\color{black} The quasipotential landscape viewpoint might provide a useful way for describing the multistability of the climate in a hierarchical fashion. 
%is flexible with respect to possible generalizations. In particular, it could accommodate multistable behavior of the climate system in a range of spatial scales, that appear in a hierarchical fashion. 
%
We present in Fig.~\ref{fig:multiscale}(a) an illustration of this perspective, where the possible states of the climate are described by the vector $X$. The quasi-potential $\Phi$ features troughs, saddles, and ridges at different scales. %The metastable states of the climate system correspond to the $\Phi$'s local minima. Random forcings allows the system to escape the deterministic basins of attraction and explore different regions of the phase space.
	
	\begin{figure}[ht!]
		\centering		
		\includegraphics[width=0.49\linewidth]{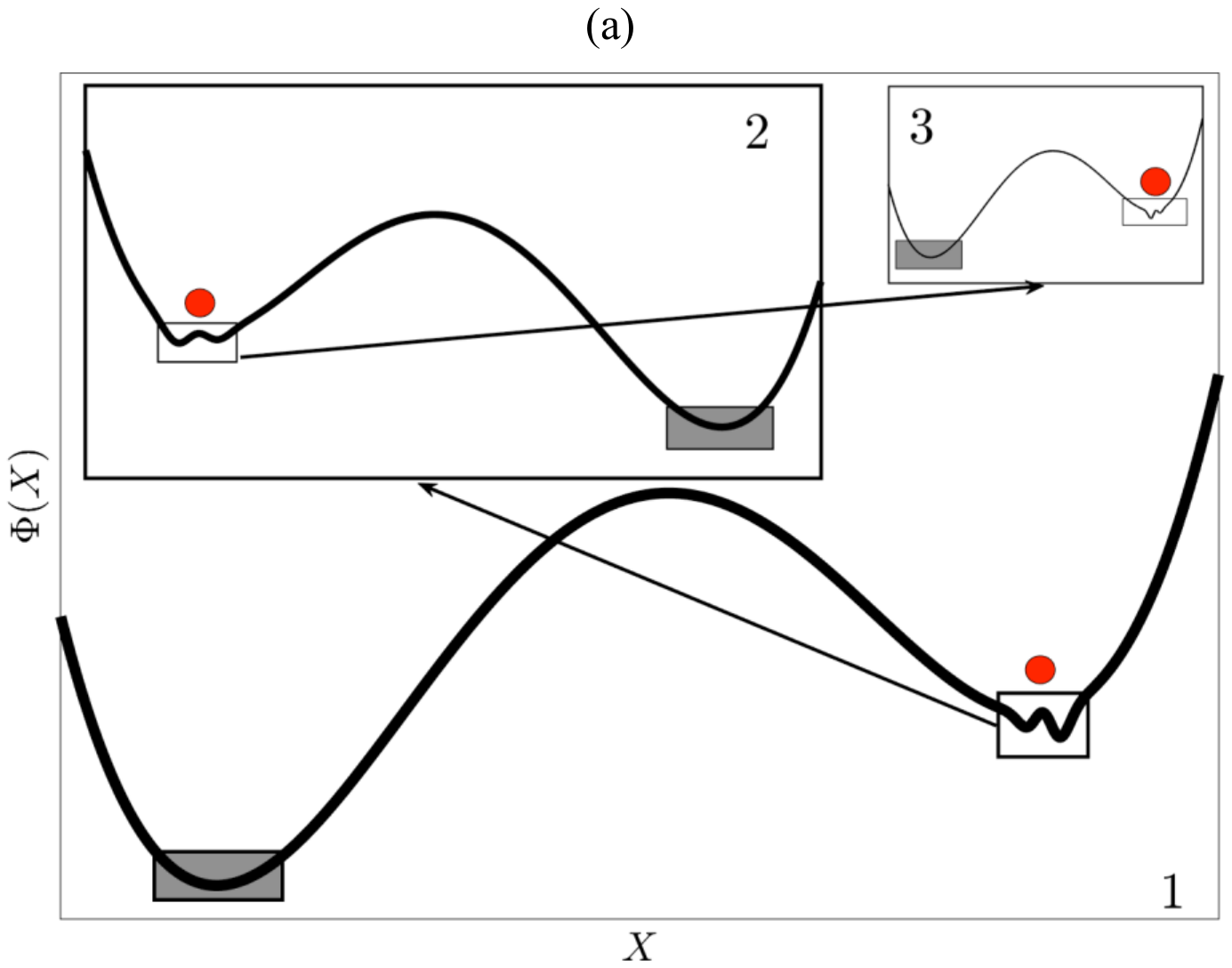}
		\includegraphics[width=0.49\linewidth]{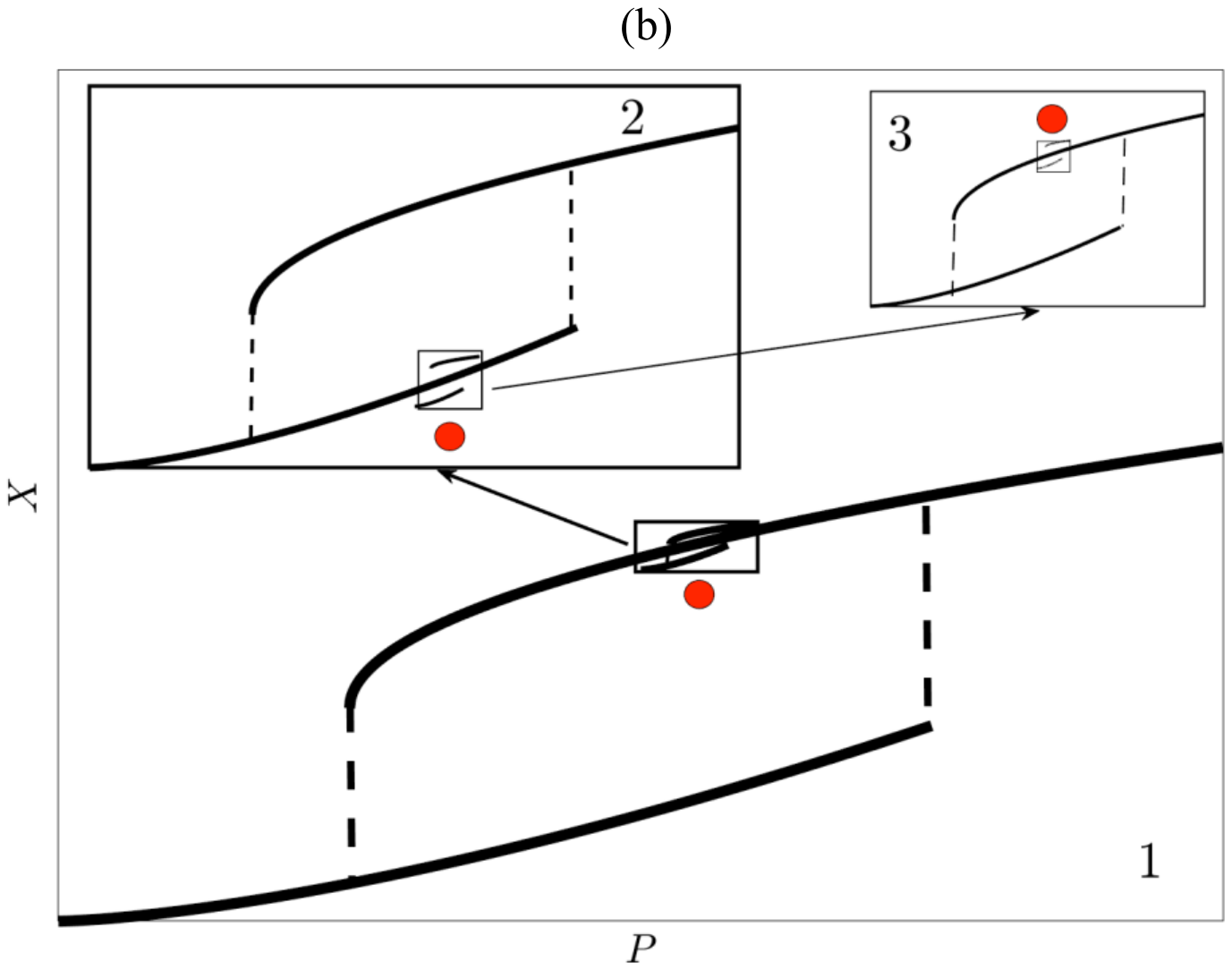}
		\caption{Schematic representation of the multiscale nature of multistability in the climate system. (a) Quasipotential $\Phi$ as a function of the state of the system $X$. (b) Corresponding hysteresis loops as a function of a parameter $P$. The white boxes indicate the zoomed-in current state of the system (red dot), going from 1 to 3 towards smaller and smaller scales.}
		\label{fig:multiscale} 
	\end{figure}
	%Let's assume, for clarity of exposition, that we consider $X$ as uni-dimensional and we identify it with the globally averaged surface temperature, where the current state of the climate is given by the red dot. This implies the fact that currently we are in an interglacial period (level 3; Holocene) of an ice-age (level 2; Pleistocene/Quaternary glaciation) taking place during a W climate state (level 1). In this metaphor, the two competing coarse-grained states differ by O(100)~K at level 1; by O(10)~K at level 2; by O(1)~K at level 3. 
	The intensity of noise allowing for exploring transitions between competing states decreases dramatically as we go from level 1 to level 3, because the local minima become shallower. Going to even smaller  scales, one would find additional (shallower) corrugations of $\Phi$. %, corresponding to the regional tipping points mentioned earlier. 
	Multistability in the climate system is often revealed by the presence of hysteresis loops obtained when suitable parameters of the system are changed, usually quasi-adiabatically \cite{Rahmstorf2005,Janssen2008,Lucarini2010}. Figure~\ref{fig:multiscale}(b) shows schematically how the multistability portrayed in Fig.~\ref{fig:multiscale}(a) appears when applying suitable protocols of parametric modulations to the system. 
	
	The above description could potentially be a fundamental mathematical structure linking the global multistability of the climate system with the geographically localized tipping elements and the so-called cascading tipping points, and might be useful for understanding the associated multiscale hysteretic behaviour of the climate system when parameters are suitably modulated. We stress that in the current work we have been able to explore only the highest hierarchical level of multistability. }
		A more complete model and a suitable, different choice of stochastic forcing would be needed for exploring the small scale local minima of the quasipotential associated, e.g., with competing climate states that exchange stability at tipping points like the ocean associated with the AMOC shutdown. In this case, one would need a model able to resolve explicitly the deep ocean circulation and possibly consider random perturbations to the hydrological cycle acting in the North Atlantic sector.

	We envision the combination of the top-down and bottom-up approach as a possible way forward to study the multiscale nature of the multistability of the climate system, as well as of other systems of comparable complexity. This research work paves the way for further investigation into some fundamental properties of the climate system and goes in the direction of clarifying its intransitive vs quasitransitive vs transitive nature \cite{Lorenz1968} when different time scales are considered. Additionally, it indicates a way for fostering the development of climate models of different level of complexity: indeed, we want them to be able to capture the qualitative features of climate, by allowing for the presence of a complex dynamical landscape featuring hierarchically arranged -- according to the desired level of envisaged detail and granularity -- competing metastable states, associated with the ensuing tipping points.  

%We foresee the possibility of interpreting and analysing the climate variability on ultralong, paleoclimatic time scales as itinerancy between possibly many competing metastable states, with proximity to M states defining times when the climate experienced sudden climatic shifts. Some metastable states may be too shallow to be practically relevant compared to the intensity of the acting stochastic and/or deterministic forcing. Other metastable states might be just very far in the phase space and have never or will very likely never be visited. 
%Transitions between metastables states might be facilitated by mechanisms like stochastic resonance \cite{Gammaitoni1998}, which has been recently reframed  according to the formalism adopted here for treating nonequilibrium systems \cite{Lucarini2019SR}. Indeed, stochastic resonance  is thought to act in the climate system at different spatial and temporal scales, ranging from ultralong \cite{Benzi1981,Nicolis1982,Ditlevsen2010}, to intermediate \cite{Alley2001,Ganopolski2002,VelezBelchi2001,Lucarini2012}, to short ones \cite{Han2014,ZhengLin2014,Wang2016}.

The viewpoint presented here seems also promising for investigating a separate, extremely relevant aspect of  atmospheric dynamics, namely the existence in the  atmosphere of different regimes of operation, which define the presence of substantial low-frequency variability on subseasonal time scales \cite{Ghil2001d,Ghil2020}. This boils down to the fact that, at coarse-grained level, due to extreme dynamical heterogeneity \cite{LucariniG2020}, one is practically looking at a multistable system, where one can define and detect transitions between different metastable states \cite{Bouchet2014}. 

Finally, we remark that white Gaussian noise might not necessarily be the only suitable way to treat stochasticity in the climate system \cite{Ditlevsen1999}. The theory of escapes from attractors in the presence of L\'evy noise has been developed \cite{Imkeller2006,Debussche2013} and very recently applied to simple geophysical  models \cite{Yang2020}. It is well known that the mechanisms of escape are rather different than in the standard Gaussian scenario pursued in this paper. It seems then of great relevance to consider the effect of L\'evy noise forcing in a more complex climate model like the one considered here. 
\nocite{Risken1996}

\enlargethispage{20pt}

\dataccess{The  data required to generate the figures can be accessed via this repository:  \href{https://doi.org/10.6084/m9.figshare.13079489}{here}. Animations presenting examples of transitions are publicly available on the \texttt{youtube.com} platform through the  links that can be found in the text that can be found at \href{https://doi.org/10.6084/m9.figshare.13079489}{here}.}

\aucontribute{GM performed the simulations, contributed to the data analysis and to the writing of the paper. TG contributed to the writing of the paper. AL contributed to the data analysis and to the writing of the paper. VL proposed the research topic, contributed to the interpretation of the data analysis and led the writing of the paper. All authors gave final approval for publication and agree to be held accountable for the work performed therein.}

\competing{The authors declare no competing interests.}

\funding{TG acknowledges the support received from the EPSRC project EP/T011866/1. VL acknowledges the support received from the EPSRC project EP/T018178/1. VL and GM acknowledge the support received from the EU Horizon 2020 project TiPES (Grant no. 820970). }

\ack{
	VL wishes to thank T. B\'odai, N. Boers, M. Ghil, F. Lunkeit, G. Pavliotis, A. Tantet, and N. Zagli for many inspiring conversations on multistability and tipping points. GM wishes to thank F. Lunkeit for his guidance on PLASIM and kind hospitality at the University of Hamburg. 
}

%\disclaimer{Insert disclaimer text here.}

%%%%%%%%%% Insert bibliography here %%%%%%%%%%%%%%

\bibliographystyle{RS}
%\bibliography{Bibliography}

\begin{thebibliography}{99}

\bibitem{Peixoto1992}
Peixoto JP, Oort AH. 1992 {\em Physics of Climate}.
New York: AIP Press, New York.

\bibitem{Lucarini2014}
Lucarini V, Blender R, Herbert C, Ragone F, Pascale S, Wouters J. 2014a
  Mathematical and physical ideas for climate science. {\em Rev. Geophys.}
  \textbf{52}, 809--859.

\bibitem{Ghil2015}
Ghil M. 2015  A mathematical theory of climate sensitivity or, {How to deal
  with both anthropogenic forcing and natural variability?}. In P. CC, M. G, M.
  L, M. WJ, editors, {\em {Climate Change : Multidecadal and Beyond}} pp.
  31--51. World Scientific Publishing Co./Imperial College Press.

\bibitem{Ghil2020}
Ghil M, Lucarini V. 2020  The physics of climate variability and climate
  change. {\em Rev. Mod. Phys.} \textbf{92}, 035002.

\bibitem{schneider1974}
Schneider SH, Dickinson RE. 1974  Climate modeling. {\em Reviews of Geophysics}
  \textbf{12}, 447--493.

\bibitem{Saltzman2001}
Saltzman B. 2001 {\em Dynamical Paleoclimatology: Generalized Theory of Global
  Climate Change}.
New York: Academic Press New York.

\bibitem{Held2005}
Held IM. 2005  The gap between simulation and understanding in climate
  modeling. {\em Bulletin of the American Meteorological Society} \textbf{86},
  1609--1614.

\bibitem{Lucarini2013}
Lucarini V. 2013  {Modeling complexity: the case of climate science}. In Gohde
  U, Hartmann S, Wolf J, editors, {\em Models, Simulations, and the Reduction
  of Complexity} pp. 229--254. De Gruyter.

\bibitem{Budyko1969}
Budyko MI. 1969  The effect of solar radiation variations on the climate of the
  Earth. {\em Tellus} \textbf{21}, 611--619.

\bibitem{sellers1969}
Sellers WD. 1969  A global climatic model based on the energy balance of the
  earth-atmosphere system. {\em Journal of Applied Meteorology} \textbf{8},
  392--400.

\bibitem{Ghil1976}
Ghil M. 1976  {Climate Stability for a Sellers-Type Model}. {\em Journal of the
  Atmospheric Sciences} \textbf{33}, 3--20.

\bibitem{Stommel1961}
Stommel H. 1961  Thermohaline convection with two stable regimes of flow. {\em
  Tellus} \textbf{2}, 244--230.

\bibitem{Veronis1963}
Veronis G. 1963  {An analysis of the wind-driven ocean circulation with a
  limited number of Fourier components}. {\em Journal of Atmospheric Sciences}
  \textbf{20}, 577--593.

\bibitem{Rooth1982}
Rooth C. 1982  Hydrology and ocean circulation. {\em Progress in Oceanography}
  \textbf{11}, 131--149.

\bibitem{Charney1979}
Charney JG, DeVore JG. 1979  Multiple flow equilibria in the atmosphere and
  blocking. {\em J. Atmos. Sci.} \textbf{36}, 1205--1216.

\bibitem{Lorenz1984}
Lorenz EN. 1984  {Irregularity: a Fundamental Property of the Atmosphere}. {\em
  Tellus A: Dynamic Meteorology and Oceanography} \textbf{36}, 98--110.

\bibitem{Lorenz1996}
Lorenz EN. 1996  {Predictability - a problem partly solved}. In Palmer T,
  Hagedorn R, editors, {\em {Predictability of Weather and Climate}} pp.
  40--58. Cambridge University Press.

\bibitem{Marshall1993}
Marshall J, Molteni F. 1993  {Toward a Dynamical Understanding of
  Planetary-Scale Flow Regimes}. {\em Journal of the Atmospheric Sciences}
  \textbf{50}, 1792--1818.

\bibitem{fraedrich1998}
Fraedrich K, Kirk E, Lunkeit F. 1998  Portable university model of the
  atmosphere.. {\em Technical Report 16 Deutsches Klimarechenzentrum}.

\bibitem{Petoukhov2000}
Petoukhov V, Ganopolski A, Brovkin V, Claussen M, Eliseev A, Kubatzki C,
  Rahmstorf S. 2000  CLIMBER-2: a climate system model of intermediate
  complexity. Part I: model description and performance for present climate.
  {\em Climate Dynamics} \textbf{16}, 1--17.

\bibitem{Montoya2006}
Montoya M, Griesel A, Levermann A, Mignot J, Hofmann M, Ganopolski A, Rahmstorf
  S. 2005  The earth system model of intermediate complexity CLIMBER-3$\alpha$.
  Part I: description and performance for present-day conditions. {\em Climate
  Dynamics} \textbf{25}, 237--263.

\bibitem{IPCC2013}
Stocker TF, Qin D, Plattner GK, Tignor M, Allen SK, Boschung J, Nauels A, Xia
  Y, Bex V, Midgley PM et~al.. 2013  Climate change 2013: The physical science
  basis. {\em Contribution of working group I to the fifth assessment report of
  the intergovernmental panel on climate change} \textbf{1535}.

\bibitem{Berner2017}
Berner J, Achatz U, Batté L, Bengtsson L, Cámara Adl, Christensen HM,
  Colangeli M, Coleman DRB, Crommelin D, Dolaptchiev SI, Franzke CLE,
  Friederichs P, Imkeller P, Järvinen H, Juricke S, Kitsios V, Lott F,
  Lucarini V, Mahajan S, Palmer TN, Penland C, Sakradzija M, von Storch JS,
  Weisheimer A, Weniger M, Williams PD, Yano JI. 2017  Stochastic
  Parameterization: Toward a New View of Weather and Climate Models. {\em
  Bulletin of the American Meteorological Society} \textbf{98}, 565--588.

\bibitem{Pierrehumbert2011}
Pierrehumbert R, Abbot D, Voigt A, Koll D. 2011  Climate of the Neoproterozoic.
  {\em Annual Review of Earth and Planetary Sciences} \textbf{39}, 417--460.

\bibitem{Hoffman1998}
Hoffman PF, Kaufman AJ, Halverson GP, Schrag DP. 1998  A Neoproterozoic
  Snowball Earth. {\em Science} \textbf{281}, 1342--1346.

\bibitem{Lewis2007}
Lewis JP, Weaver AJ, Eby M. 2007  Snowball versus slushball Earth: Dynamic
  versus nondynamic sea ice?. {\em Journal of Geophysical Research: Oceans}
  \textbf{112}.

\bibitem{Lucarini2010}
Lucarini V, Fraedrich K, Lunkeit F. 2010  Thermodynamic analysis of snowball
  Earth hysteresis experiment: Efficiency, entropy production and
  irreversibility. {\em Quarterly Journal of the Royal Meteorological Society}
  \textbf{136}, 2--11.

\bibitem{Abbot2011}
Abbot DS, Voigt A, Koll D. 2011  The Jormungand global climate state and
  implications for Neoproterozoic glaciations. {\em Journal of Geophysical
  Research: Atmospheres} \textbf{116}.

\bibitem{Lucarini2017b}
Lucarini V, B{\'{o}}dai T. 2017  Edge states in the climate system: exploring
  global instabilities and critical transitions. {\em Nonlinearity}
  \textbf{30}, R32--R66.

\bibitem{Lenton2008}
Lenton TM, Held H, Kriegler E, Hall JW, Lucht W, Rahmstorf S, Schellnhuber HJ.
  2008  Tipping elements in the Earth's climate system. {\em Proceedings of the
  national Academy of Sciences} \textbf{105}, 1786--1793.

\bibitem{Boers2017}
Boers N, Marwan N, Barbosa HMJ, Kurths J. 2017  A deforestation-induced tipping
  point for the South American monsoon system. {\em Scientific Reports}
  \textbf{7}, 41489.

\bibitem{Rahmstorf2005}
Rahmstorf S, Crucifix M, Ganopolski A, Goosse H, Kamenkovich I, Knutti R,
  Lohmann G, Marsh R, Mysak LA, Wang Z, Weaver AJ. 2005  Thermohaline
  circulation hysteresis: A model intercomparison. {\em Geophysical Research
  Letters} \textbf{32}.

\bibitem{Walter2006}
Walter KM, Zimov SA, Chanton JP, Verbyla D, Chapin FS. 2006  Methane bubbling
  from Siberian thaw lakes as a positive feedback to climate warming. {\em
  Nature} \textbf{443}, 71--75.

\bibitem{Levermann2009}
Levermann A, Schewe J, Petoukhov V, Held H. 2009  Basic mechanism for abrupt
  monsoon transitions. {\em Proceedings of the National Academy of Sciences}
  \textbf{106}, 20572--20577.

\bibitem{Steffen2018}
Steffen W, Rockstr{\"o}m J, Richardson K, Lenton TM, Folke C, Liverman D,
  Summerhayes CP, Barnosky AD, Cornell SE, Crucifix M, Donges JF, Fetzer I,
  Lade SJ, Scheffer M, Winkelmann R, Schellnhuber HJ. 2018  Trajectories of the
  Earth System in the Anthropocene. {\em Proceedings of the National Academy of
  Sciences} \textbf{115}, 8252--8259.

\bibitem{Klose2020}
Klose AK, Karle V, Winkelmann R, Donges JF. 2020  Emergence of cascading
  dynamics in interacting tipping elements of ecology and climate. {\em Royal
  Society Open Science} \textbf{7}, 200599.

\bibitem{Gammaitoni1998}
Gammaitoni L, H\"{a}nggi P, Jung P, Marchesoni F. 1998  {Stochastic resonance}.
  {\em Reviews of Modern Physics} \textbf{70}, 223--287.

\bibitem{Lucarini2019SR}
Lucarini V. 2019  Stochastic resonance for nonequilibrium systems. {\em Phys.
  Rev. E} \textbf{100}, 062124.

\bibitem{Benzi1981}
Benzi R, Sutera A, Vulpiani A. 1981  The mechanism of stochastic resonance.
  {\em Journal of Physics A: Mathematical and General} \textbf{14}, L453--L457.

\bibitem{Nicolis1982}
Nicolis C. 1982  Stochastic aspects of climatic transitions‚Äìresponse to a
  periodic forcing. {\em Tellus} \textbf{34}, 308--308.

\bibitem{Ditlevsen2010}
Ditlevsen PD. 2010  Extension of stochastic resonance in the dynamics of ice
  ages. {\em Chemical Physics} \textbf{375}, 403 -- 409.

\bibitem{Alley2001}
Alley RB, Anandakrishnan S, Jung P. 2001  Stochastic resonance in the North
  Atlantic. {\em Paleoceanography} \textbf{16}, 190--198.

\bibitem{Ganopolski2002}
Ganopolski A, Rahmstorf S. 2002  Abrupt Glacial Climate Changes due to
  Stochastic Resonance. {\em Phys. Rev. Lett.} \textbf{88}, 038501.

\bibitem{VelezBelchi2001}
V\'{e}lez-Belch\'{i} P, Alvarez A, Colet P, Tintore J, Haney RL. 2001
  Stochastic resonance in the thermohaline circulation.. {\em Geophysical
  Research Letters} \textbf{28}, 2053--2056.

\bibitem{Lucarini2012}
Lucarini V, Faranda D, Willeit M. 2012  Bistable systems with stochastic noise:
  virtues and limits of effective one-dimensional Langevin equations. {\em
  Nonlinear Processes in Geophysics} \textbf{19}, 9--22.

\bibitem{Han2014}
Han Q, Yang T, Zeng C, Wang H, Liu Z, Fu Y, Zhang C, Tian D. 2014  {Impact of
  time delays on stochastic resonance in an ecological system describing
  vegetation}. {\em Physica A: Statistical Mechanics and its Applications}
  \textbf{408}, 96--105.

\bibitem{ZhengLin2014}
Jia ZL, Yang CY, Li C, Mei DC. 2014  Stochastic resonance in a
  groundwater-dependent plant ecosystem with fluctuations and time delay. {\em
  Journal of Statistical Mechanics: Theory and Experiment} \textbf{2014}.

\bibitem{Wang2016}
Wang KK, Wang YJ, Wu JC. 2016  Stochastic resonance and stability for an
  ecological vegetation growth system driven by colored noises and
  multiplicative signal. {\em Modern Physics Letters B} \textbf{30}, 1650308.

\bibitem{Lucarini2020}
Lucarini V, B{\'{o}}dai T. 2020  Global stability properties of the climate:
  Melancholia states, invariant measures, and phase transitions. {\em
  Nonlinearity} \textbf{33}, R59--R92.

\bibitem{Waddington1957}
Waddington C, Kacser H. 1957 {\em The Strategy of the Genes: A Discussion of
  Some Aspects of Theoretical Biology}.
Allen \& Unwin.

\bibitem{Ao2009}
Ao P. 2009  Global view of bionetwork dynamics: adaptive landscape. {\em
  Journal of Genetics and Genomics} \textbf{36}, 63 -- 73.

\bibitem{Ferrell2012}
Ferrell JE. 2012  Bistability, bifurcations, and Waddington’s epigenetic
  landscape. {\em Curr. Biol.} \textbf{22}, R458.

\bibitem{Huang2012}
Huang S. 2012  The molecular and mathematical basis of Waddington’s
  epigenetic landscape: A framework for post-Darwinian biology?. {\em
  BioEssays} \textbf{34}, 149.

\bibitem{Lucarini2019}
Lucarini V, B\'odai T. 2019  Transitions across Melancholia States in a Climate
  Model: Reconciling the Deterministic and Stochastic Points of View. {\em
  Phys. Rev. Lett.} \textbf{122}, 158701.

\bibitem{fraedrich2005}
Fraedrich K, Jansen H, Kirk E, Luksch U, Lunkeit F. 2005  The Planet Simulator:
  Towards a user friendly model. {\em Meteorologische Zeitschrift} \textbf{14},
  299--304.

\bibitem{Lucarini2010b}
Lucarini V, Fraedrich K, Lunkeit F. 2010  Thermodynamics of climate change:
  generalized sensitivities. {\em Atmospheric Chemistry and Physics}
  \textbf{10}, 9729--9737.

\bibitem{Boschi2013}
Boschi R, Lucarini V, Pascale S. 2013  Bistability of the climate around the
  habitable zone: A thermodynamic investigation. {\em Icarus} \textbf{226},
  1724 -- 1742.

\bibitem{Holden2016}
Holden PB, Edwards N, Fraedrich K, Kirk E, Lunkeit F, Zhu X. 2016  PLASIM-GENIE
  v1.0: a new intermediate complexity AOGCM. {\em Geoscientific Model
  Development} \textbf{9}, 3347--3361.

\bibitem{ragone2016new}
Ragone F, Lucarini V, Lunkeit F. 2016  A new framework for climate sensitivity
  and prediction: a modelling perspective. {\em Climate Dynamics} \textbf{46},
  1459--1471.

\bibitem{Lucarini2017a}
Lucarini V, Ragone F, Lunkeit F. 2017  Predicting climate change using response
  theory: Global averages and spatial patterns. {\em Journal of Statistical
  Physics} \textbf{166}, 1036--1064.

\bibitem{Holden2019}
Holden PB, Edwards NR, Rangel TF, Pereira EB, Tran GT, Wilkinson RD. 2019
  PALEO-PGEM v1.0: a statistical emulator of Pliocene--Pleistocene climate.
  {\em Geoscientific Model Development} \textbf{12}, 5137--5155.

\bibitem{Graham1987}
Graham R. 1987  Macroscopic potentials, bifurcations and noise in dissipative
  systems. In {\em Fluctuations and Stochastic Phenomena in Condensed Matter}
  pp. 1--34. Springer.

\bibitem{Hamm1994}
Hamm A, Tél T, Graham R. 1994  Noise-induced attractor explosions near tangent
  bifurcations. {\em Physics Letters A} \textbf{185}, 313 -- 320.

\bibitem{Zhou2012}
Zhou JX, Aliyu MDS, Aurell E, Huang S. 2012  Quasi-potential landscape in
  complex multi-stable systems. {\em Journal of The Royal Society Interface}
  \textbf{9}, 3539--3553.

\bibitem{Zhou2016}
Zhou P, Li T. 2016  Construction of the landscape for multi-stable systems:
  Potential landscape, quasi-potential, A-type integral and beyond. {\em The
  Journal of Chemical Physics} \textbf{144}, 094109.

\bibitem{faccoID}
Facco E, d'Errico M, Rodriguez A, Laio A. 2017  Estimating the intrinsic
  dimension of datasets by a minimal neighborhood information. {\em Scientific
  reports} \textbf{7}, 12140.

\bibitem{rodriguez2018computing}
Rodriguez A, d'Errico M, Facco E, Laio A. 2018  Computing the Free Energy
  without Collective Variables. {\em Journal of chemical theory and
  computation} \textbf{14}, 1206--1215.

\bibitem{automatic-topography}
d'Errico M, Facco E, Laio A, Rodriguez A. 2018  Automatic topography of
  high-dimensional data sets by non-parametric Density Peak clustering. {\em
  arXiv preprint arXiv:1802.10549}.

\bibitem{science_DPA}
Rodriguez A, Laio A. 2014  Clustering by fast search and find of density peaks.
  {\em Science} \textbf{344}, 1492--1496.

\bibitem{Grebogi1983}
Grebogi C, Ott E, Yorke JA. 1983  Fractal Basin Boundaries, Long-Lived Chaotic
  Transients, and Unstable-Unstable Pair Bifurcation. {\em Phys. Rev. Lett.}
  \textbf{50}, 935--938.

\bibitem{Vollmer2009}
Vollmer J, Schneider TM, Eckhardt B. 2009  Basin boundary, edge of chaos and
  edge state in a two-dimensional model. {\em New Journal of Physics}
  \textbf{11}, 013040.

\bibitem{LT:2011}
Lai YC, T{\'e}l T. 2011 {\em Transient chaos: complex dynamics on finite time
  scales} vol. 173.
Springer Science \& Business Media.

\bibitem{Skufca2006}
Skufca JD, Yorke JA, Eckhardt B. 2006  Edge of Chaos in a Parallel Shear Flow.
  {\em Physical Review Letters} \textbf{96}, 174101.

\bibitem{bodai2015}
B{\'o}dai T, Lucarini V, Lunkeit F, Boschi R. 2015  Global instability in the
  Ghil--Sellers model. {\em Climate Dynamics} \textbf{44}, 3361--3381.

\bibitem{hanggi1986}
H\"anggi P. 1986  Escape from a metastable state. {\em Journal of Statistical
  Physics} \textbf{42}, 105--148.

\bibitem{Grassberger1989}
Grassberger P. 1989  Noise-induced escape from attractors. {\em Journal of
  Physics A: Mathematical and General} \textbf{22}, 3283--3290.

\bibitem{freidlin1998}
Freidlin MI, Wentzell AD. 1984 {\em Random perturbations of dynamical systems}.
New York: Springer.

\bibitem{Bell2004}
Bell DR. 2004  Stochastic differential equations and hypoelliptic operators. In
  {\em Real and stochastic analysis} pp. 9--42. Springer.

\bibitem{Graham1991}
Graham R, Hamm A, T\'el T. 1991  Nonequilibrium potentials for dynamical
  systems with fractal attractors or repellers. {\em Phys. Rev. Lett.}
  \textbf{66}, 3089--3092.

\bibitem{Gaspard2002}
Gaspard P. 2002  Trace formula for noisy flows. {\em Journal of statistical
  physics} \textbf{106}, 57--96.

\bibitem{Bouchet2016}
Bouchet F, Gawedzki K, Nardini C. 2016  Perturbative calculation of
  quasi-potential in non-equilibrium diffusions: a mean-field example. {\em
  Journal of Statistical Physics} \textbf{163}, 1157--1210.

\bibitem{Ao2004}
Ao P. 2004  Potential in stochastic differential equations: novel construction.
  {\em Journal of Physics A: Mathematical and General} \textbf{37}, L25--L30.

\bibitem{Yuan2017}
Yuan R, Tang Y, Ao P. 2017  SDE decomposition and A-type stochastic
  interpretation in nonequilibrium processes. {\em Frontiers of Physics}
  \textbf{12}, 120201.

\bibitem{Kautz1987}
Kautz R. 1987  Activation energy for thermally induced escape from a basin of
  attraction. {\em Physics Letters A} \textbf{125}, 315 -- 319.

\bibitem{Grafke2015}
Grafke T, Grauer R, Sch{\"a}fer T. 2015  The instanton method and its numerical
  implementation in fluid mechanics. {\em Journal of Physics A: Mathematical
  and Theoretical} \textbf{48}, 333001.

\bibitem{Grafke2017}
Grafke T, Sch\"afer T, Vanden-Eijnden E. 2017  Long {Term} {Effects} of {Small}
  {Random} {Perturbations} on {Dynamical} {Systems}: {Theoretical} and
  {Computational} {Tools}. In {\em Recent {Progress} and {Modern} {Challenges}
  in {Applied} {Mathematics}, {Modeling} and {Computational} {Science}} Fields
  {Institute} {Communications} pp. 17--55. Springer, New York, NY.

\bibitem{Grafke2019}
Grafke T, Vanden-Eijnden E. 2019  Numerical computation of rare events via
  large deviation theory. {\em Chaos: An Interdisciplinary Journal of Nonlinear
  Science} \textbf{29}, 063118.

\bibitem{Kramers1940}
Kramers HA. 1940  Brownian motion in a field of force and the diffusion model
  of chemical reactions. {\em Physica} \textbf{7}, 284.

\bibitem{fractal_dim}
Grassberger P, Procaccia I. 2004  Measuring the strangeness of strange
  attractors. In {\em The Theory of Chaotic Attractors} pp. 170--189. Springer.

\bibitem{Frederickson1983}
Frederickson P, Kaplan JL, Yorke ED, Yorke JA. 1983  The liapunov dimension of
  strange attractors. {\em Journal of Differential Equations} \textbf{49}, 185
  -- 207.

\bibitem{Lucarini2016Extremes}
Lucarini V, Faranda D, de~Freitas ACGMM, de~Freitas JMM, Holland M, Kuna T,
  Nicol M, Todd M, Vaienti S. 2016 {\em Extremes and Recurrence in Dynamical
  Systems}.
New York: Wiley.

\bibitem{Faranda2017}
Faranda D, Messori G, Yiou P. 2017  Dynamical proxies of North Atlantic
  predictability and extremes. {\em Scientific Reports} \textbf{7}, 41278.

\bibitem{knn}
Mack Y, Rosenblatt M. 1979  Multivariate k-nearest neighbor density estimates.
  {\em Journal of Multivariate Analysis} \textbf{9}, 1--15.

\bibitem{Eyring2016}
Eyring V, Bony S, Meehl GA, Senior CA, Stevens B, Stouffer RJ, Taylor KE. 2016
  {Overview of the Coupled Model Intercomparison Project Phase 6 (CMIP6)
  experimental design and organization}. {\em Geoscientific Model Development}
  \textbf{9}, 10539--10583.

\bibitem{Orszag1970}
Orszag SA. 1970  {Transform Method for the Calculation of Vector-Coupled Sums:
  Application to the Spectral Form of the Vorticity Equation}. {\em Journal of
  the Atmospheric Sciences} \textbf{27}, 890--895.

\bibitem{haltiner1980}
Haltiner GJ, Williams RT. 1982 {\em Numerical prediction and dynamic
  meteorology}.
New York: John Wiley and Sons.

\bibitem{Lembo2020}
Lembo V, Lucarini V, Ragone F. 2020  Beyond Forcing Scenarios: Predicting
  Climate Change through Response Operators in a Coupled General Circulation
  Model. {\em Scientific Reports} \textbf{10}, 8668.

\bibitem{Semtner1976}
Semtner, Albert~J. J. 1976  {A Model for the Thermodynamic Growth of Sea Ice in
  Numerical Investigations of Climate}. {\em Journal of Physical Oceanography}
  \textbf{6}, 379--389.

\bibitem{Dijkstra2005}
Dijkstra HA, Ghil M. 2005  Low-frequency variability of the large-scale ocean
  circulation: A dynamical systems approach. {\em Reviews of Geophysics}
  \textbf{43}, RG3002.

\bibitem{Kuhlbrodt2007}
Kuhlbrodt T, Griesel A, Montoya M, Levermann A, Hofmann M, Rahmstorf S. 2007
  On the driving processes of the Atlantic meridional overturning circulation.
  {\em Reviews of Geophysics} \textbf{45}.

\bibitem{Cessi2019}
Cessi P. 2019  The global overturning circulation. {\em Annual Review of Marine
  Science} \textbf{11}, 249--270.

\bibitem{Trenberth2009}
Trenberth KE, Fasullo JT, Kiehl J. 2009  Earth's Global Energy Budget. {\em
  Bulletin of the American Meteorological Society} \textbf{90}, 311--324.

\bibitem{Lucarini2011}
Lucarini V, Ragone F. 2011  Energetics of Climate Models: Net Energy Balance
  and Meridional Enthalpy Transport. {\em Rev. Geophys.} \textbf{49}, RG1001.

\bibitem{Knietzsch2015}
Knietzsch MA, Schr\"oder A, Lucarini V, Lunkeit F. 2015  The impact of oceanic
  heat transport on the atmospheric circulation. {\em Earth System Dynamics}
  \textbf{6}, 591--615.

\bibitem{lorenz1967}
Lorenz EN, Lorenz FN. 1967 {\em The nature and theory of the general
  circulation of the atmosphere} vol. 218.
World Meteorological Organization Geneva.

\bibitem{Weiss2020}
Weiss JB, Fox-Kemper B, Mandal D, Nelson AD, Zia RKP. 2020  Nonequilibrium
  Oscillations, Probability Angular Momentum, and the Climate System. {\em
  Journal of Statistical Physics} \textbf{179}, 1010--1027.

\bibitem{Pauluis2002}
Pauluis O, Held IM. 2002  Entropy budget of an atmosphere in
  radiative-convective equilibrium. {P}art {II}: {L}atent heat transport and
  moist processes. {\em J. Atmos. Sci.} \textbf{59}, 140--149.

\bibitem{Lucarini2009}
Lucarini V. 2009  {Thermodynamic efficiency and entropy production in the
  climate system}. {\em Physical Review E} \textbf{80}, 021118.

\bibitem{Goody2000}
Goody R. 2000  Sources and sinks of climate entropy. {\em Q. J. R. Meteorol.
  Soc.} \textbf{126}, 1953--1970.

\bibitem{Lucarini2014a}
Lucarini V, Pascale S. 2014  {Entropy production and coarse graining of the
  climate fields in a general circulation model}. {\em Climate Dynamics}
  \textbf{43}, 981--1000.

\bibitem{Gaspard2004}
Gaspard P. 2004  Time-Reversed Dynamical Entropy and Irreversibility in
  Markovian Random Processes. {\em Journal of Statistical Physics}
  \textbf{117}, 599--615.

\bibitem{Linsenmeier2015}
Linsenmeier M, Pascale S, Lucarini V. 2015  Climate of Earth-like planets with
  high obliquity and eccentric orbits: Implications for habitability
  conditions. {\em Planetary and Space Science} \textbf{105}, 43 -- 59.

\bibitem{Hyde2000}
Hyde WT, Crowley TJ, Baum SK, Peltier WR. 2000  Neoproterozoic `snowball
  Earth'simulations with a coupled climate/ice-sheet model. {\em Nature}
  \textbf{405}, 425--429.

\bibitem{schutte2011markov}
Sch{\"u}tte C, No{\'e} F, Lu J, Sarich M, Vanden-Eijnden E. 2011  Markov state
  models based on milestoning. {\em The Journal of chemical physics}
  \textbf{134}, 05B609.

\bibitem{Rockstr2009}
Rockstr{\"o}m J, Steffen W, Noone K, Persson {\AA}, Chapin FS, Lambin EF,
  Lenton TM, Scheffer M, Folke C, Schellnhuber HJ, Nykvist B, de~Wit CA, Hughes
  T, van~der Leeuw S, Rodhe H, S{\"o}rlin S, Snyder PK, Costanza R, Svedin U,
  Falkenmark M, Karlberg L, Corell RW, Fabry VJ, Hansen J, Walker B, Liverman
  D, Richardson K, Crutzen P, Foley JA. 2009  A safe operating space for
  humanity. {\em Nature} \textbf{461}, 472--475.

\bibitem{Janssen2008}
Janssen RHH, Meinders MBJ, Van~Nes EH, Scheffer M. 2008  Microscale
  vegetation-soil feedback boosts hysteresis in a regional vegetation–climate
  system. {\em Global Change Biology} \textbf{14}, 1104--1112.

\bibitem{Lorenz1968}
Lorenz EN. 1968  Climatic Determinism. In Mitchell JM, editor, {\em Causes of
  Climatic Change} pp. 1--13. American Meteorological Society, Boston.

\bibitem{Ghil2001d}
Ghil M, Robertson AW. 2002  "Waves" vs. "particles" in the atmosphere's phase
  space: {A pathway to long-range forecasting?}. {\em Proc. Natl. Acad. Sci.
  USA} \textbf{99}, (Suppl. 1), 2493--2500.

\bibitem{LucariniG2020}
Lucarini V, Gritsun A. 2020  A new mathematical framework for atmospheric
  blocking events. {\em Climate Dynamics} \textbf{54}, 575--598.

\bibitem{Bouchet2014}
Bouchet F, Laurie J, Zaboronski O. 2014  Langevin Dynamics, Large Deviations
  and Instantons for the Quasi-Geostrophic Model and Two-Dimensional Euler
  Equations. {\em Journal of Statistical Physics} \textbf{156}, 1066--1092.

\bibitem{Ditlevsen1999}
Ditlevsen PD. 1999  Observation of $alpha$-stable noise induced millennial
  climate changes from an ice-core record. {\em Geophysical Research Letters}
  \textbf{26}, 1441--1444.

\bibitem{Imkeller2006}
Imkeller P, Pavlyukevich I. 2006  First exit times of SDEs driven by stable
  L\'evy processes. {\em Stochastic Processes and their Applications}
  \textbf{116}, 611 -- 642.

\bibitem{Debussche2013}
Debussche A, H{\"o}gele M, Imkeller P. 2013 {\em The Dynamics of Nonlinear
  Reaction-Diffusion Equations with Small L{\'e}vy Noise}.
Lecture Notes in Mathematics. Springer International Publishing.

\bibitem{Yang2020}
Yang F, Zheng Y, Duan J, Fu L, Wiggins S. 2020  The tipping times in an Arctic
  sea ice system under influence of extreme events. {\em Chaos: An
  Interdisciplinary Journal of Nonlinear Science} \textbf{30}, 063125.

\bibitem{Risken1996}
Risken H. 1996 {\em The Fokker-Planck equation}.
Springer, Berlin.

\end{thebibliography}

\vskip2pc

\end{document}